# The Impact of Modern AI in Metadata Management

Wenli Yang[1] · Rui Fu[1] · Muhammad Bilal Amin[1] · Byeong Kang[1]

**Abstract**
Metadata management plays a critical role in data governance, resource discovery, and decision-making in the data-driven era. While traditional metadata approaches have primarily focused on organization, classification, and resource reuse, the integration of modern artificial intelligence (AI) technologies has significantly transformed these processes. This paper investigates both traditional and AI-driven metadata approaches by examining open-source solutions, commercial tools, and research initiatives. A comparative analysis of traditional and AI-driven metadata management methods is provided, highlighting existing challenges and their impact on next-generation datasets. The paper also presents an innovative AI-assisted metadata management framework designed to address these challenges. This framework leverages more advanced modern AI technologies to automate metadata generation, enhance governance, and improve the accessibility and usability of modern datasets. Finally, the paper outlines future directions for research and development, proposing opportunities to further advance metadata management in the context of AI-driven innovation and complex datasets.

## Abbreviations

| | |
|---|---|
| AI | Artificial Intelligence |
| GPT | Generative Pretrained Transformer |
| MMS | Metadata Management System |
| NLP | Natural Language Processing |
| GDPR | General Data Protection Regulation |
| HIPAA | Health Insurance Portability and Accountability Act |
| ISO | International Organization for Standardization |
| LLM | Large language model |
| OCR | Optical Character Recognition |
| AWS | Amazon Web Services |
| ECS | Elastic Container Service |
| HTTP | HyperText Transfer Protocol |
| API | Application Programming Interface |
| YAML | YAML Ain't Markup Language |
| RDF | Resource Description Framework |
| JDBC | Java Database Connectivity |
| PHP | Hypertext Preprocessor |
| SQLite | Structured Query Language Database |
| DCAT | Data Catalog Vocabulary |
| DCMI | Dublin Core Metadata Initiative |
| OData | Open Data Protocol |
| DHS | DataHub Service |
| ODD | OpenDataDiscovery |
| NLU | Natural Language Understanding |
| IoT | Internet of Things |
| JSON | JavaScript Object Notation |
| XML | Extensible Markup Language |
| CSV | Comma-Separated Values |
| ML | Machine Learning |
| DL | Deep Learning |
| GenAI | Generative AI |
| GNNs | Graph Neural Networks |
| KGs | Knowledge Graphs |
| RAI | Responsible AI |
| XAI | Explainable AI |

## 1 Introduction

Metadata management is essential across various fields, as it provides detailed descriptions necessary for interpreting and utilizing data effectively. From research and personal information management to industries like digital archives, metadata offers context, records transactions, and establishes relationships between datasets [1]. Metadata standards further support traceability, ensuring data accuracy, integrity,

✉ Wenli Yang
yang.wenli@utas.edu.au

1 School of ICT, University of Tasmania, Tasmania, Australia

and trustworthiness, while promoting resource discovery and reuse [2, 3]. Libraries and organizations increasingly rely on metadata to organize, classify, and inventory data efficiently, underscoring its growing importance [3].

In the age of AI, metadata management has become even more critical. The rapid expansion of data demands concise and precise metadata to streamline the discovery of relevant information and improve AI training processes. Metadata enables effective data curation, ensuring high-quality, well-contextualized datasets that enhance AI performance and decision-making capabilities. Efficient metadata lifecycle management—from creation to archiving—is vital for maintaining relevance, improving retrieval efficiency, and supporting accountability systems in today's data-driven landscape. As metadata evolves, it continues to play a pivotal role in shaping AI-driven innovations, improving data governance, and enhancing the value of audiovisual and other resources [4, 5].

The increasing complexity and volume of data require advanced management techniques, particularly in domains like healthcare, where big data analytics and metadata are crucial for improving decision-making and preventive strategies [5, 6]. Proper metadata practices bridge the semantic gap between data and user understanding, enabling efficient data governance, integration, and quality assurance. Metadata also supports digital transformation by allowing users to discover and reuse data effectively [7, 8]. AI technologies, such as natural language processing (NLP) and unsupervised learning, further enhance metadata generation and maintenance by automating the extraction of relevant information and addressing challenges like data quality and governance [1, 8].

Despite advancements, integrating AI into metadata management faces challenges, including ensuring practical effectiveness and addressing the lack of in-depth research on AI applications in this area. Current literature highlights AI's role in healthcare and information systems but provides limited insights into metadata management practices [9, 10]. Future research must explore specific methods for applying AI to metadata management, focusing on issues like metadata reuse and lifecycle management. Emerging technologies, such as large language models (LLMs) like generative pretrained transformer 4 (GPT-4), offer opportunities to revolutionize metadata systems by improving efficiency and accuracy [11, 12].

This research aims to analyse mainstream metadata management systems, both open-source and proprietary, to evaluate their use of AI-enhanced functional modules. It investigates how AI technologies are integrated into these systems to improve performance and explores new directions for future applications. By addressing practical issues like metadata reuse and lifecycle management, this study seeks to provide actionable AI-driven solutions and lay the groundwork for next-generation metadata management systems [13, 14]. This survey paper provides the following contributions:

- **Provide a comprehensive description of the evolution of metadata management**: This paper details the transition from traditional metadata management approaches to AI-driven innovations, offering an in-depth analysis of both traditional and AI-based solutions in the context of metadata management.
- **Identify and analyse gaps and challenges in current metadata management approaches**: The paper highlights existing gaps, including unresolved issues in traditional methods and limitations in current AI-powered tools. It also explores the impact of these gaps on next-generation datasets, underlining the necessity of addressing these challenges.
- **Proposed AI-Assisted Metadata Solution for Future Datasets:** proposes a novel AI-assisted solution for metadata management that aims to address existing gaps. It includes a high-level architecture diagram and outlines the key features and capabilities of the proposed system. The contribution focuses on how the solution will improve metadata handling, enhance data quality, and impact the management of future datasets.

## 2 Overview of Metadata Management

This section provides a comprehensive overview of the key components of metadata management, starting with the metadata lifecycle, which describes the stages data undergoes throughout its existence. Next, the different types of metadata will be explored, highlighting their roles in providing context and structure to data. Finally, we will examine the management structure of metadata, focusing on common use of meta data across diverse systems. Understanding these foundational elements is essential for leveraging metadata in the context of modern data-driven applications and AI technologies.

### 2.1 Metadata Lifecycle

The metadata lifecycle typically consists of four main stages: acquisition, cleaning, verification, and maintenance. Each of these stages plays a crucial role in ensuring that metadata remains accurate, reliable, and valuable throughout its entire lifecycle. Figure 1 illustrates the metadata lifecycle process, describing the flow of data through each of these stages, highlighting key activities and their importance in maintaining metadata quality and consistency.

**Acquisition**: Data acquisition refers to the process of gathering existing data [15]. This step involves collecting metadata from various sources, such as indexes, abstracts, and bibliographic records. The data is often created based on cataloguing rules and standards, ensuring consistency across different datasets.

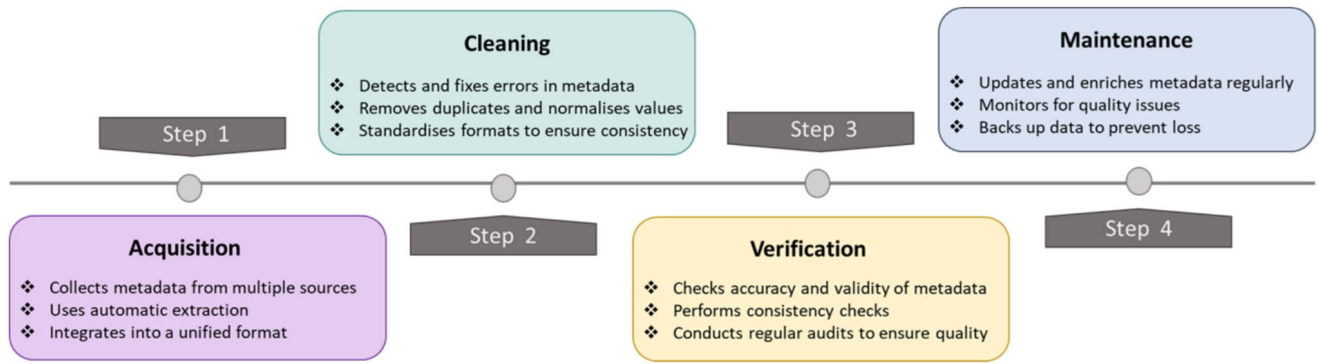

**Fig. 1** Metadata lifecycle process: stages and key activities

**Cleaning**: Cleaning metadata involves verifying the integrity and accuracy of metadata records [16]. This process includes ensuring that metadata fields are fully populated, as well as identifying and correcting any errors or inconsistencies. Effective cleaning ensures that the metadata is of high quality and ready for use in various applications.

**Verification**: Metadata verification focuses on confirming the accuracy and completeness of metadata records [17]. This step typically involves conducting sample checks to ensure that the metadata meets relevant standards and conforms to established requirements, guaranteeing its reliability.

**Maintenance**: Continuous maintenance of metadata is essential for preserving its quality and ensuring its ongoing effectiveness [18]. This involves regularly updating metadata records, implementing backup and recovery procedures, and instituting security measures to protect metadata from unauthorized access or tampering.

Each of these stages ensures that metadata remains a valuable asset for data discovery, analysis, and decision-making across systems and applications.

### 2.2 Types of Metadata

Metadata originates from multiple sources and file types for storage, including drug management, patient status management, geographic information, image management, enterprise management, and social media user profiles. The file types for storing metadata encompass formats like Comma-Separated Values (CSV), Excel, JavaScript Object Notation (JSON), Extensible Markup Language (XML) and various database files. The Table 1 provides a comprehensive overview of the different types of metadata, detailing their definitions, examples, purposes, common use cases, and associated file types.

**Descriptive Metadata**: This type of metadata provides details about the content of a resource, such as its title, author, subject, and keywords. It helps users locate and understand the resource. It is commonly used in digital libraries, academic databases, and search engines.

**Structural Metadata**: Structural metadata describes the internal structure of a resource, such as its table of contents, pagination, chapters, and sections, and is frequently found in eBooks, websites, and multimedia resources. It aids users in navigating and understanding the organization of the resource.

**Administrative Metadata**: Administrative metadata offers information about the creation and management of a resource, such as file format, creation date, copyright information, and storage location. It plays a crucial role in archiving, document management, and content management systems.

**Technical Metadata**: This metadata describes the technical characteristics of a resource, such as resolution, file size, and software used, which is essential for resource identification, processing, and preservation in fields like image/video processing and digital asset management.

**Use Metadata**: Use metadata records information about how a resource is accessed or utilized, such as user ratings, access logs, and usage statistics, helping measure the impact of resources, especially in web analytics, user feedback, and e-commerce platforms.

### 2.3 Metadata Management System Architecture and Modules

The metadata management system (MMS) is composed of several critical modules that work in unison to ensure the efficient storage, retrieval, and governance of metadata. Figure 2 Illustrates the six key modules of a Metadata Management System. These modules serve distinct yet interconnected roles in the overall structure of the system, supporting key functions such as metadata capture, classification and storage. In this section, we will explore the essential components of a Metadata Management System, highlighting their individual functions and how they contribute to the seamless operation of the system.

**Table 1** Overview of different types of metadata

| Metadata type | Examples | Purpose | Common use cases | Typical file types | Refs |
| --- | --- | --- | --- | --- | --- |
| Descriptive metadata | Title, Author, Keywords, Subject, Abstract, ISBN | Helps in discovery and identification of resources | Cataloguing, Search Engines, Academic Databases | CSV, Excel, JSON, XML | [19, 20] |
| Structural metadata | Table of Contents, Pagination, Chapters, Sections, File Organization | Assists in organizing and navigating resources | eBooks, Websites, Multimedia Resources, Digital Archives | JSON, XML, HTML, Databases | [21, 22] |
| Administrative metadata | File Format, Creation Date, Copyright, Storage Location, Access Rights | Ensures resource management, preservation, and access control | Archiving, Document Management, Content Management Systems | XML, JSON, Databases | [23, 24] |
| Technical metadata | File Size, Resolution, Bit Depth, Compression, Software Used | Assists in resource identification, processing, and preservation | Image/Video Processing, Digital Asset Management, Software Archives | TIFF, PNG, JPEG, MP4, DOCX | [25, 26] |
| Use metadata | Access Logs, User Ratings, Review Count, Annotations, Usage Statistics | Measures resource impact, tracks interactions, and improves access | Web Analytics, User Feedback, Digital Media Libraries, E-commerce | CSV, JSON, Databases | [27, 28] |

**Metadata Extraction and Generation**: This module is fundamental for building a metadata ecosystem by identifying and extracting metadata from diverse data sources, including structured databases, semi-structured files like XML and JSON, and unstructured formats such as text documents and multimedia files. It employs advanced automated tools and machine learning (ML), deep learning (DL) algorithms to analyse the properties and content of data, ensuring metadata generation is thorough and consistent. Key features include the ability to handle complex datasets, support for multiple file formats, and integration with external data catalogues for enrichment. This module is critical for maintaining an accurate and comprehensive reflection of the underlying data's structure, attributes, and interrelationships, facilitating downstream metadata processes.

**Metadata Search and Discovery**: This module enables users to locate and retrieve metadata across large and complex datasets with ease. It leverages cutting-edge technologies such as NLP, semantic search, and intelligent filtering techniques to enhance search precision and usability. Users can perform keyword searches, browse metadata by predefined categories, or use advanced options like contextual or faceted searches. In addition to supporting efficient data discovery, this module incorporates personalized suggestions, search history tracking, and visual tools like metadata maps or dashboards to provide insights into metadata patterns. By improving data accessibility and usability, this module empowers users to derive value from metadata more effectively.

**Metadata Quality Management**: Ensuring the trustworthiness and usability of metadata is the primary objective of this module. It includes rigorous validation mechanisms to check metadata against predefined rules and standards, identifying errors, inconsistencies, or redundancies. Automated error detection and resolution tools enhance the reliability of metadata, while data cleaning processes refine its accuracy and consistency. The module also assesses metadata completeness and compliance with organizational or regulatory requirements. Additional capabilities such as metadata enrichment—adding context or supplementary information—further enhance its quality. By maintaining high-quality metadata, this module supports better data-driven decision-making and minimizes the risk of errors in metadata utilization.

**Metadata Storage and Indexing**: This module serves as the backbone of metadata management by providing scalable and secure storage solutions. It ensures metadata is stored efficiently in centralized repositories, distributed systems, or hybrid environments, depending on the organization's needs. Advanced indexing mechanisms organize metadata for rapid retrieval, enabling real-time access even in large-scale systems. Security features such as encryption, role-based access control, and audit logs protect sensitive metadata from unauthorized access. The module also supports disaster recovery through backup and failover systems, ensuring metadata integrity and availability. Its scalability ensures that as data volumes grow, the metadata repository can expand seamlessly.

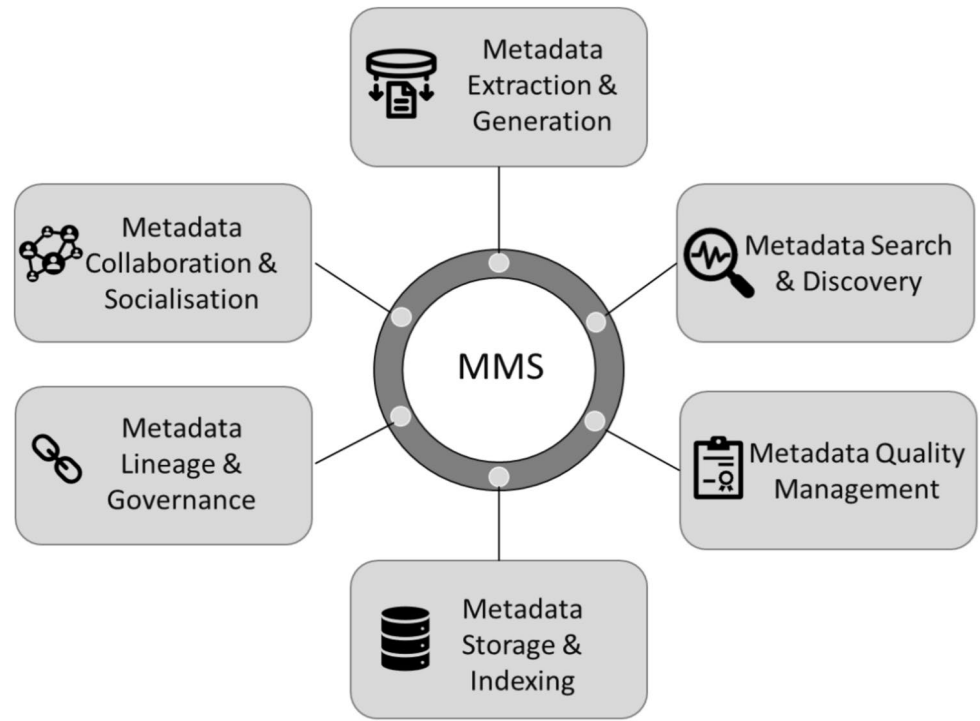

**Fig. 2** Key modules of metadata management system

**Metadata Lineage and Governance**: Transparency, accountability, and compliance are the hallmarks of this module. Metadata lineage tracks the origins, flow, and transformations of metadata throughout its lifecycle, creating a comprehensive audit trail. This ensures that changes to metadata are transparent and can be traced back to their source. Governance frameworks within the module establish clear policies and standards for metadata management, including rules for data access, usage, and retention. Compliance features ensure adherence to regulatory requirements such as General Data Protection Regulation (GDPR), Health Insurance Portability and Accountability Act (HIPAA), or International Organization for Standardization (ISO) standards. By providing clear oversight and control, this module enhances organizational trust in metadata systems and promotes informed decision-making.

**Metadata Collaboration and Socialization**: This module transforms metadata management into a collaborative process by facilitating user interaction and collective knowledge sharing. It provides tools for annotating, tagging, and commenting on metadata, enabling teams to add context and insights. Rating systems and user feedback mechanisms help prioritize and refine metadata based on relevance or accuracy. The module also integrates with communication platforms like Slack or Microsoft Teams, streamlining collaborative workflows. Version control and change tracking ensure transparency in updates and modifications. By fostering a culture of collaboration, this module enhances user engagement and democratizes metadata management, making it more accessible and impactful for diverse stakeholders.

Table 2 provides a comprehensive overview of the key modules in metadata management systems, highlighting their core functions, the AI techniques employed, the typical application domains.

## 3 Evolution of Metadata Management: From Traditional to AI-Driven Innovations

Metadata management has evolved significantly over time, adapting to the growing complexity of data systems. In its early stages, traditional approaches to metadata management focused on manual processes and structured methods for organizing and maintaining data. These methods were often labour-intensive and required a high level of human intervention. However, the rise of AI has led to the development of more automated and intelligent solutions. This shift has spurred the creation of both open-source and commercial tools, alongside increasing research efforts aimed at improving metadata management with AI. This section outlines the progression from traditional to AI-driven metadata management approaches.

**Table 2** Key AI techniques and applications in metadata management modules

| Module | Key functions | AI techniques used | Typical application domain | Refs |
| --- | --- | --- | --- | --- |
| Metadata extraction &generation | – Extract metadata from diverse data sources<br>– Analyse data properties and content<br>– Generate comprehensive metadata | ML/DL, NLP, Generative AI (GenAI) | Data warehouses, Internet of Things (IoT) systems, Healthcare | [29–31] |
| Metadata search & discovery | – Locate metadata across repositories<br>– Perform contextual and semantic searches<br>– Personalize search results | NLP, Semantic Search, Graph Neural Networks (GNNs), Knowledge Graphs (KGs) | Research, E-commerce, social media | [32] |
| Metadata quality management | – Validate and clean metadata<br>– Ensure consistency and accuracy<br>– Enrich metadata with supplementary info | Rule-based AI, Anomaly Detection, ML/DL, NLP, Reinforcement Learning | Finance, Scientific Research | [33] |
| Metadata storage & indexing | – Store metadata securely and efficiently<br>– Enable fast metadata retrieval<br>– Ensure scalability and integrity | AI-powered Indexing, Predictive Modelling, NLP | Cloud computing, Logistics, Retail | [20, 34] |
| Metadata lineage & governance | – Track metadata flow and transformations<br>– Ensure compliance with policies<br>– Define governance standards | AI for Compliance, Provenance Tracking Algorithms | Legal, Pharmaceuticals, Smart Cities | [35] |
| Metadata collaboration & socialization | – Enable sharing and annotation of metadata<br>– Support feedback and version control | Recommendation Systems, Sentiment Analysis, Chatbots, LLMs | Education, Business Intelligence, Media | [36–38] |

### 3.1 Traditional Metadata Approaches

This section examines four prominent metadata management tools: Amundsen, CKAN, Meta-Grid, and Atlas. These tools represent a variety of traditional approaches to metadata management, catering to different organizational needs. Amundsen, CKAN, and Meta-Grid are open-source tools, while Atlas is a commercial solution. By exploring their design philosophies, key features, deployment options, and customization capabilities, this study aims to provide insights into the landscape of traditional metadata management tools, which do not incorporate AI techniques.

The selection of these tools is based on their distinct approaches to metadata management. Some tools focus on automating metadata curation and integration, while others provide frameworks for dataset description and comprehensive data inventory management. These tools, despite their differences, play a critical role in improving data governance, transparency, and collaboration within organizations. Furthermore, the research highlights ongoing efforts to improve metadata management, offering a foundation for future developments in the field.

#### 3.1.1 Open-Source Solutions for Metadata

Open-source solutions have become popular for metadata management due to their cost-effectiveness, flexibility, and the ability to be customized for specific organizational needs.

**Amundsen**[1]: is an open-source data catalogue designed to improve the productivity of data analysts, scientists, and engineers by leveraging automated and curated metadata. It integrates with popular data resources and databases, such as Apache Cassandra and PostgreSQL, facilitating seamless interaction with tables, people, and dashboards. Deployment options for Amundsen include Amazon Web Services (AWS) Elastic Container Service (ECS), Docker with Gunicorn, and Apache open-source deployment strategies. Key modules in Amundsen include a data ingestion library and a metadata application programming interface (API) that manage metadata efficiently, covering table descriptions, column details, tags, and lineage information. Although it does not utilize AI techniques, Amundsen offers broad integration capabilities with various data platforms and cloud services. Comprehensive documentation is provided, and community support is available via email and online forms. Notably, Amundsen is a cost-free solution with no licensing fees, but customization options are limited.

---

[1] Amundsen Metadata Service. 2020; Available from: https://www.amundsen.io/amundsen/metadata/

**CKAN**[2]: offers a robust MMS that provides a framework for dataset description in formats such as CSV, Excel, and JSON. Customization of metadata schemas is supported through extensions, such as YAML Ain't markup language (YAML) or JSON schema descriptions, which allow tailored validation and template snippets. CKAN provides APIs for metadata interaction, including the FileStore API for file uploads and the CKAN API for accessing dataset metadata via hyperText transfer protocol (HTTP) POST requests. Metadata-rich datasets are equipped with intuitive labeling, which facilitates search, sharing, and linking. CKAN's extensible metadata management supports geospatial data, versioning, and integration with other catalogs through plugins like data catalog vocabulary (DCAT). The connection of publications with datasets enhances data context. Customization is enabled through CKAN's IDatasetForm plugin interface for custom metadata fields, the resource description framework (RDF) Serializer for dataset metadata profiles, and YAML or JSON schema descriptions for schema configuration and sharing.

**Meta-Grid**[3]: is a comprehensive data catalog tool designed for organizations of all sizes. It provides a centralized platform for data inventory and documentation management, available as both a self-hosted solution and a cloud-based variant, Meta-Grid Cloud. The tool emphasizes data democratization, allowing all users to view data content, while only authorized personnel can modify or delete it, supported by granular rights management that enables custom role configurations. Meta-Grid comprises several key components: an installer and updater tool developed in Python and Java (LiquiBase), a web-based user interface built with hypertext preprocessor (PHP) (Yii2) and hosted on an Apache Webserver, a relational database storage system using structured query language database (SQLite) (with plans to support additional database backends), and a bulk loader for importing metadata from various Java database connectivity (JDBC) sources via Java (Pentaho Data Integration). While Meta-Grid is a robust solution for data management, the current implementation does not incorporate AI technologies. As such, it remains a traditional solution for metadata management, without the AI-driven features like intelligent data discovery or automated data quality monitoring.

### 3.1.2 Commercial Metadata Tools

Commercial metadata management tools such as Atlas provide enterprise-level solutions designed to address the complex needs of large organizations. These tools offer comprehensive functionalities, including data discovery, metadata curation, and data governance, often combined with advanced security features and seamless integration with other enterprise applications.

Atlan[4] is a prominent commercial tool for metadata management, known for its ability to automate metadata curation and facilitate data integration. It supports various data types, making it suitable for organizations that work with diverse data sources. Atlan also provides robust data governance tools, enabling organizations to control access, usage, and ensure compliance with legal and regulatory standards. This makes Atlan particularly valuable for industries such as healthcare, finance, and government, where compliance and security are critical.

A key advantage of commercial tools like Atlan is the level of support they offer. These tools typically come with dedicated customer service teams, training resources, and regular updates to help organizations maximize the software's potential. For large organizations with complex data infrastructures, commercial tools provide high availability, reliability, and technical support, which may not be as readily available with open-source alternatives.

Commercial tools generally offer more advanced features than open-source solutions. Atlan, for example, includes integrated data lineage tracking, allowing organizations to visualize how data flows through different systems and transformations. This feature is essential for maintaining data quality and accountability, particularly in regulated industries where the origin and transformations of data must be clearly understood for compliance purposes.

Furthermore, commercial metadata tools like Atlan integrate seamlessly with other enterprise software systems, such as business intelligence tools, data warehouses, and cloud platforms. This integration helps organizations

---

[2] CKAN. 2024; Available from: https://ckan.org/.
[3] Meta-Grid. 2023; Available from: https://github.com/patschwork/meta_grid
[4] Atlan. 2024; Available from: https://atlan.com/?ref=/about/

centralize their data management efforts and gain a unified view of their data assets.

### 3.1.3 Research Initiatives in Metadata Management

During the non-AI periods, research initiatives in metadata management primarily focused on addressing key challenges related to large-scale, distributed, and diverse data environments. One major area of focus was the automation of metadata curation, a task that had traditionally required manual tagging, classification, and updates. These manual processes were often time-consuming, error-prone, and inefficient. To overcome these challenges, researchers explored methods to improve metadata management processes through more automated solutions. Although ML and NLP were not prevalent at this time, scholars investigated simpler computational approaches to streamline the tagging and classification of data, enhancing the overall efficiency of metadata management systems.

Another important area of research involved metadata standardization and interoperability. With the increasing variety of data sources across multiple platforms, ensuring consistent metadata across these systems became a significant concern. As a response, research efforts were directed towards the development of standardized metadata schemas aimed at promoting interoperability and facilitating integration between disparate data systems. Initiatives such as the Dublin Core Metadata Initiative (DCMI) [39] and the Open Data Protocol (OData) [40] played a crucial role in establishing common standards for metadata management, fostering data sharing and integration across platforms.

Additionally, metadata played a critical role in data discovery, which was essential for helping users locate and comprehend relevant datasets [41]. Improved metadata search capabilities and enhanced tagging methodologies were developed, which led to the creation of more advanced tools for indexing and retrieving metadata. These advancements helped make datasets more accessible, promoting better data discovery and usability for researchers and analysts alike.

## 3.2 AI-Driven Metadata Approaches

The contemporary landscape of data management is increasingly shaped by the widespread adoption of AI technologies, including NLP, ML algorithms, and GenAI models. These technologies are transforming metadata management by automating processes and enhancing capabilities in data discovery, cataloguing, and governance. As AI becomes more integrated into data management systems, it is challenging traditional assumptions and enabling the automation of metadata management across diverse platforms, driving significant advancements in the field.

### 3.2.1 AI-Driven Open-Source Platforms

This subsection examines notable AI-driven open-source tools, including DataGalaxy, DataHub, Magda, OpenDataDiscovery (ODD), and OpenMetadata. These platforms integrate advanced functionalities, such as NLP for data classification, clustering algorithms, and innovative techniques for metadata management. Through automation of metadata mapping, extraction, and organization, these tools enhance data cataloging, governance, and compliance. By improving data observability and accessibility, they support streamlined metadata management processes for contemporary organizations.

**DataGalaxy**[5]: emphasizes advanced metadata management by utilizing AI-powered features for data classification and organization. It integrates NLP to automate data classification and supports clustering algorithms for efficient data organization. The platform streamlines metadata workflows and enables real-time collaboration and governance through its intuitive data catalogue. Additionally, DataGalaxy offers AI-driven suggestions and multilingual capabilities, making it easier for teams to manage and share data insights. Its robust metadata management solutions enhance data discovery and compliance while facilitating improved decision-making and data governance practices.

**DataHub**[6]: recognized as a third-generation metadata platform, supports data discovery, collaboration, governance, and end-to-end observability. It accommodates various file types, including schemeless formats like CSV, JSONL, and JSON. Deployment options include Kubernetes clusters, local open-source installations, and cloud-based solutions via the DataHub Service (DHS). Key components of DataHub include the ingestion framework for modular data ingestion, the metadata service for metadata operations, and the schema registry for schema management. The platform also features plugins for integrating diverse data sources. AI-driven functionalities enhance data governance, quality, and analytics, while APIs enable interaction with the Metadata Graph. DataHub supports pre-built integrations with platforms such as Kafka, Apache Atlas and multiple databases [42, 43], with additional capabilities for custom applications via its API.

**Magda**[7]: is an open-source data catalogue platform designed to streamline the management of data assets and metadata. Magda enhances data organization and accessibility by automating the extraction, classification, and linking of metadata. By utilizing AI techniques, it offers federated data discovery, allowing users to search datasets from various sources without moving the data. Additionally, Magda's metadata enhancement features improve the quality of dataset descriptions by automatically deriving metadata without transmitting the underlying data. It also supports integration with various data sources and provides robust tools for tracking changes, detecting duplications, and ensuring secure data access through customizable authorization systems. These features collectively ensure efficient and transparent data governance.

---

[5] DataGalaxy. 2024; Available from: https://www.datagalaxy.com/en/

[6] DataHub. 2024; Available from: https://datahubproject.io/

[7] Magda. 2024; Available from: https://magda.io/

**Table 3** Key AI features integrated into current open-source AI metadata tools

| Tool | Key AI techniques used |
| --- | --- |
| DataGalaxy | NLP for automating data classification and enhancing data insights. Clustering for efficient cataloging. Recommendation systems based on data context |
| DataHub | AI for Data Discovery and Analytics: Uses a Modular Ingestion Framework to streamline data ingestion |
| Magda | Metadata Extraction using AI to automatically extract and organize metadata. Federated Data Discovery for searching distributed data without movement. AI-driven Metadata Enhancement to improve dataset descriptions and organization |
| OpenDataDiscovery | AI for Categorization automates data categorization. Lineage Tracking using AI to track data flow and dependencies |
| OpenMetadata | NLP for automating tagging and organizing sensitive data like PII. AI for Metadata Organization streamlines metadata classification and compliance tasks |

**OpenDataDiscovery (ODD)**[8]**:** offers a comprehensive open-source platform designed to enhance data governance. It integrates key functionalities such as data discovery, quality management, glossary creation, lineage tracking, data modeling, and security measures. By centralizing these elements, ODD provides a unified view of metadata, helping organizations ensure data accessibility, compliance, and transparency. Leveraging AI techniques, ODD automates data classification, lineage tracing, and quality assessment, improving governance and enabling seamless integration with other data tools. This holistic approach supports accurate, reliable, and accessible metadata for better decision-making and regulatory compliance.

**OpenMetadata**[9]**:** is an open-source, unified platform designed to facilitate data discovery, observability, and governance across an organization. OpenMetadata enhances metadata management by utilizing AI-driven capabilities to simplify the documentation process [44]. By applying NLP during data ingestion, it automates the identification and tagging of sensitive personally identifiable information (PII). This functionality not only facilitates the efficient organization of metadata but also ensures compliance with data privacy regulations. OpenMetadata strengthens data governance, allowing organizations to maintain secure, accurate, and compliant metadata, which is essential for informed decision-making and regulatory adherence.

Table 3 summarizes the key AI features integrated into current commercial AI metadata tools.

### 3.2.2 Commercial AI Metadata Tools

This section contrasts closed-source tools such as Alation, Ataccama ONE Data Catalog, Atlan Data Catalog, Azure Data Catalog, Collibra, Datafold Data Catalog, Informatica, Monte Carlo, Select Star, and Talend. These proprietary platforms, using the latest versions, leverage AI technologies, including NLP, ML, and GenAI, to automate key data management processes. These include data discovery, metadata crawling, classification, lineage tracing, and governance. By integrating AI insights with domain expertise, these tools enhance data quality, security, and reliability across various data ecosystems, facilitating advanced functionalities like context-based search and automated documentation.

**Alation**[10]**:** employs NLP to enhance named entity recognition and improve recommendation systems. The platform includes ALLIE AI, which supports data administrators in organizing data during ingestion. Bidirectional interfaces further strengthen data discovery, lineage tracking, and governance. Combining ML algorithms with domain expertise, Alation automates the classification, labeling, and identification of sensitive information, ensuring better data management.

**Ataccama**[11]**:** utilizes AI to extract metadata from a variety of data sources. It integrates GenAI to facilitate self-service data discovery, cataloging data from databases and AI-driven systems. Its Data Catalog also leverages a wide range of AI technologies, including ML and NLP, to automate metadata management. It supports data discovery, classification, profiling, lineage tracking, and self-service functionalities.

**Atlan**[12]**:** latest version of Atlan also provides comprehensive metadata management, organizing and maintaining metadata to enhance data discovery, governance, and access. The platform supports bulk uploads and offers flexible deployment options, including on-premises and cloud-based solutions with compatibility for AWS, Azure, and Google Cloud Platform. Atlan integrates ML and NLP to automate data discovery and metadata management, unifying metadata from multiple sources into a single platform. It also offers APIs for accessing metadata and tracking data movement and integrates with tools like DBT for expanded capabilities.

---

[8] OpenDataDiscovery. 2024; Available from: https://opendatadiscovery.org/

[9] OpenMetadata. 2024; Available from: https://open-metadata.org/

[10] Alation. 2024; Available from: https://www.alation.com/

[11] Ataccama. 2024; Available from: https://www.ataccama.com/

[12] Atlan. 2024; Available from: https://atlan.com/?ref=/about/

Table 4  Key AI features integrated into current commercial AI metadata tools

| Tool | Key AI features |
| --- | --- |
| Alation | NLP for entity recognition, AI for data organization, data discovery, and lineage tracking |
| Ataccama | AI for metadata extraction, self-service discovery, GenAI for cataloguing data |
| Atlan | AI for metadata crawling, classification, lineage tracking, integrates with DBT |
| Azure Data Catalog | AI and ML for metadata annotation, NLP, computer vision, and predictive analytics |
| Collibra | Context-based search, AI-driven governance, and automated classification with GenAI models |
| Datafold | AI for data profiling, metadata management, and lineage tracing |
| Informatica | CLAIRE AI engine for automation, GenAI applications, low-code/no-code interfaces for data democratization |
| Monte Carlo | Metadata management, data cataloguing, lineage tracking, integration with data quality tools using AI |
| Select Star | AI-driven automation for metadata organization and insights into data relationships |
| Talend | AI for data discovery, profiling, metadata management, and governance compliance |

**Azure Data Catalog**[13]: leverages AI and ML to enhance data discovery and management. It incorporates AI algorithms for metadata annotation and utilizes ML models for organizing and discovering data assets. Additionally, Azure's AI services, including NLP, computer vision, and predictive analytics, support improved data management functions.

**Collibra**[14]: incorporates AI to offer context-based search and automate connections between disparate data sources. The platform enhances data governance, improves quality, and supports advanced analytics through a GenAI model trained on historical data and metadata. This facilitates the automation of processes such as data classification.

**Datafold**[15]: utilizes AI technologies, such as NLP, to automate key aspects of data management. These capabilities streamline the processes of data profiling, discovery, metadata management, and lineage tracking, enabling organizations to efficiently manage their data assets. By automating these functions, Datafold improves the accuracy and consistency of data operations, enhances traceability, and supports more effective decision-making, all while reducing the manual effort typically involved in managing complex data environments.

**Informatica**[16]: utilizes the CLAIRE AI engine to optimize enterprise data management by automating routine tasks and consolidating metadata intelligence. Key capabilities include embedding GenAI technology into data management to enhance efficiency and support digital transformation. It also democratizes data access by simplifying discovery through intuitive natural language interfaces. Additionally, CLAIRE enables users to create GenAI applications with minimal technical barriers through low-code or no-code platforms, accelerating development and deployment for a broader user base.

**Monte Carlo**[17]: is a crucial feature of its Data Observability platform. It automates data cataloging and metadata handling to ensure traceability of data lineage, which improves data quality, reliability, and security. The platform supports metadata uploads from external tables and views, with continuous collection from sources like Snowflake's information schema. Additionally, Monte Carlo offers multi-cloud hosting for metadata management and utilizes data collectors to extract metadata from data warehouses, lakes, and BI tools. GenAI is incorporated for tasks such as dashboard generation and team performance analysis. Integration with tools like Atlan and vector databases further enhances data quality assurance. The platform provides customer support and customization options, allowing users to configure and interact with metadata and insights from the Monte Carlo analytics engine, ensuring comprehensive data coverage and reliability.

**Select Star**[18]: automates the processes of data analysis and documentation. It provides insights into the organization of metadata, the relationships between data, and the usage patterns within data systems. By leveraging AI-driven processes, Select Star facilitates the efficient management and exploration of data, allowing users to easily access and interpret metadata, improving overall data governance and insights.

---

[13] Azure Data Catalog. 2024; Available from: https://azure.microsoft.com/en-au/products/data-catalog/
[14] Collibra. 2024; Available from: https://www.collibra.com/us/en
[15] Datafold. 2024; Available from: https://www.datafold.com/
[16] Informatica. 2024; Available from: https://www.informatica.com/
[17] Monte Carlo. 2024; Available from: https://www.montecarlodata.com/
[18] Select Star. 2024; Available from: https://www.selectstar.com/

**Talend**[19]: automates various stages of data management, including discovery, crawling, profiling, and organization. Using AI, it enriches and links metadata, ensuring that all data assets are properly governed. Talend's managed metadata system guarantees compliance with data governance standards while fostering internal knowledge discovery, enabling better decision-making and more efficient data management processes across organizations.

Table 4 summarizes the key AI features integrated into current commercial AI metadata tools.

### 3.2.3 Research Initiatives in AI Metadata

In recent years, there has been growing interest in AI-driven metadata management, driven by the increasing need to manage large volumes of structured and unstructured data. Research efforts focus on developing advanced metadata systems that go beyond traditional automation to improve data discovery, quality, and governance. ML, NLP, and GenAI are being employed to enhance metadata extraction, data lineage tracking, and data quality assessments. This section examines the AI techniques applied in metadata management, highlighting their advantages, challenges, and potential applications for efficient and scalable data management solutions.

We conducted both a descriptive analysis and a comparative analysis of AI techniques used in metadata management. The descriptive analysis focused on explaining how each technique is applied, what it is used for, and its main strengths and weaknesses. The comparative analysis looked at what each technique can do across a set of common functional areas to support decision-making about their use in different modules.

Table 5 provides a descriptive summary of selected AI techniques used in metadata management. Each row presents an AI technique aligned with one of the key modules of the Metadata Management System (MMS) introduced in Sect. 2.3. The table outlines the main advantages of each technique, its known limitations, and typical use cases. References are included to provide supporting evidence from existing studies. This structured overview helps illustrate how different AI techniques are applied across metadata processes and highlights the trade-offs involved in their practical use.

Furthermore, in order to evaluate and compare AI techniques used in MMS, we identify a set of core functional capabilities that reflect the typical demands of metadata processing. These dimensions represent how different techniques contribute to various metadata tasks such as content extraction, categorisation, pattern discovery, and dynamic adaptation, within the context of a structured and comprehensive way to evaluate each technique's effectiveness by focusing on functional capabilities [90, 91]. This helps identify not only how well a method performs technically, but also how useful and interpretable it is in real-world metadata management.

Table 6 presents a comparative view of these functional capabilities. It includes evaluation criteria such as extraction, classification, clustering, reasoning, real-time support, human-in-the-loop integration, explainability, scalability, and interpretability. Each capability is marked to show whether the technique strong support (✓), partial/conditional support (⚠), or does not support (✗) that function. This side-by-side comparison helps identify which techniques are best suited for specific metadata management challenges. The following descriptions clarify the meaning and scope of each capability used in the comparison.

- **Extraction:** Extraction refers to a technique's ability to automatically identify and retrieve relevant metadata or features from raw input data such as text, images, audio, or sensor logs. It focuses on transforming unstructured or semi-structured data into structured metadata that can be stored, queried, or further processed.
- **Classification:** Classification involves assigning metadata items to predefined categories or labels based on input characteristics. This is a supervised task, requiring known label sets, and is commonly used for content labelling, filtering, or organizing metadata entries into specific types or domains.
- **Clustering:** Clustering is the process of grouping similar metadata items based on feature similarity, without using predefined labels. This unsupervised approach helps discover hidden patterns or relationships within metadata, such as grouping documents by topic or records by behavioural traits.
- **Reasoning:** Reasoning enables a system to derive new metadata, relationships, or decisions by applying logical rules, learned policies, or structured relationships. This may include inferring missing metadata, validating consistency, or drawing conclusions based on hierarchical or semantic connections within existing metadata.
- **Real-Time Support:** The technique supports live data processing and responsive metadata operations. It can process streaming or incoming metadata inputs and deliver outputs immediately or with minimal delay. This enables its use in real-time metadata workflows or system loops, where timely decision-making and low-latency responses are critical.

---

[19] Talend. 2024; Available from: https://www.talend.com/

**Table 5** AI techniques in metadata management—use cases, advantages and limitations

| Modules | Used AI techniques | Advantages using AI techniques | Limitations | Use cases | Refs |
|---|---|---|---|---|---|
| Metadata extraction and generation | NLP | Allows for extracting structured metadata from unstructured text, improving the efficiency of metadata management by identifying key entities and relationships | High resource demand, customization for specific domains may be complex and costly | Text classification, named entity recognition, and document metadata extraction | [45] |
| | Rule-based Systems | Utilizes predefined rules to standardize and automate metadata extraction, improving consistency and accuracy | Rigid, hard to scale across diverse and evolving datasets, and lacks flexibility for handling novel metadata | Automated metadata tagging, document categorization, content management systems | [46] |
| | ML Clustering | Automatically identifies and groups similar metadata, helping to organize large and complex datasets into manageable clusters, aiding in more efficient metadata extraction | Requires careful parameter tuning for different datasets and may not perform well with unstructured or noisy data | Metadata grouping for content categorization, automated metadata tagging | [47, 48] |
| | ML classifier | Efficiently classifies metadata into predefined categories, making it useful for high-dimensional and complex metadata tasks | Efficiency drops with large datasets and may require substantial fine-tuning for optimal performance | Text classification, metadata tagging, document organization, content filtering | [49] |
| | Segmentation and Feature Extraction | Automates the extraction of relevant metadata features, which improves both the processing and organization of metadata | Can suffer from instability and generalization issues, especially when applied to small or heterogeneous datasets | Content metadata generation, image and video metadata tagging | [50, 51] |
| | Optical Character Recognition (OCR) | Extracts text from scanned or printed documents, converting it into editable and searchable metadata | OCR accuracy can be affected by text quality, layout, and image distortions, requiring post-processing for optimal results | Document digitization, metadata extraction from scanned documents | [52] |
| | Audio Annotation | Helps in the classification and tagging of audio data with relevant metadata, facilitating better organization and searchability | Depends heavily on expert knowledge for creating accurate annotations, limiting scalability across diverse audio types | Audio file organization, metadata tagging for podcasts, and voice recordings | [53] |
| | DL | Can automatically learn complex metadata features, making it useful for extracting valuable patterns in large-scale datasets | Requires large amounts of labelled training data, and model interpretability may be limited, leading to potential overfitting | Metadata extraction for image recognition, content tagging, and speech recognition | [54, 55] |
| | LLMs | Efficiently handles vast amounts of textual metadata by understanding language patterns, enabling automated content extraction and semantic analysis | Consumes significant computational resources, requires large-scale data for training, and may lack transparency in decision-making | Text summarization, automated document metadata generation, and NLP-based metadata processing | [56–58] |

**Table 5** (continued)

| Modules | Used AI techniques | Advantages using AI techniques | Limitations | Use cases | Refs |
| --- | --- | --- | --- | --- | --- |
| Metadata Search and Discovery | Information Retrieval | Optimizes metadata search by ranking and retrieving relevant metadata from large datasets using advanced search algorithms | Search quality heavily depends on the metadata quality and may suffer from performance issues with ambiguous queries | Metadata search engines, content discovery, document retrieval, and indexing | [59] |
| | Reinforcement Learning | Enables adaptive metadata search by continuously improving search strategies based on user interactions and feedback | Requires large amounts of interaction data and may be computationally expensive for real-time searches | Personalized search results, adaptive metadata recommendation, and dynamic discovery systems | [60] |
| | Natural Language Understanding (NLU) | Enhances metadata search by understanding query context, enabling more accurate and semantically relevant search results | Can be complex and resource-intensive to train on specific domains or languages | Semantic search engines, intelligent content discovery, and context-aware query processing | [61, 62] |
| | Collaborative Filtering | Improves metadata discovery by analysing user behaviour to recommend relevant metadata, offering personalized results based on historical data | May suffer from the "cold start" problem with limited user data or new metadata sets | Content recommendation, personalized search, and metadata discovery in large databases | [63] |

**Table 5** (continued)

| Modules | Used AI techniques | Advantages using AI techniques | Limitations | Use cases | Refs |
|---|---|---|---|---|---|
| Metadata Quality management | Anomaly Detection, ML Clustering | Identifies inconsistencies or unusual patterns in metadata, which improves data quality by detecting errors and outliers | May produce false positives or miss subtle errors, especially in noisy data | Data validation, quality monitoring, and error detection in metadata | [47, 64, 65] |
| | Rule-based Systems | Automates metadata validation and error detection by applying predefined rules or decision-making logic to metadata, improving consistency and completeness | Can be rigid and difficult to scale to diverse datasets, may require manual tuning | Ensuring completeness and consistency, cleaning metadata across datasets | [66, 67] |
| | ML Classification, DL | Automatically identifies and corrects metadata issues by learning from data patterns, making error correction more efficient and scalable | Requires large datasets for training and may suffer from overfitting if not properly regularized | Error correction, identifying faulty metadata entries, and improving quality | [68–70] |
| | Reinforcement Learning, Evolutionary Algorithms, Bayesian Optimization | Continuously optimizes and improves metadata quality by learning from feedback and iterating through generations of metadata quality improvements | Requires significant data and computational resources, may be slow to converge to optimal solutions | Adaptive quality improvement, continuous learning for metadata systems | [14, 71] |
| | NLP | Automatically fills in missing metadata or validates text-based metadata entries, ensuring completeness and accuracy of metadata | Requires domain-specific training data and computational power for complex tasks | Text-based metadata validation, ensuring data completeness and consistency | [14, 72] |
| | Edge AI, Federated Learning | Provides real-time monitoring and validation of metadata, allowing for immediate error detection and quality improvement | Requires proper hardware infrastructure for edge computing, may struggle with scalability in large systems | Real-time metadata monitoring and immediate feedback for error correction | [73] |
| | Fuzzy Logic Systems, GNNs, KGs | Analyses relationships and dependencies between metadata entities to detect errors, redundancies, and ensure that metadata is logically consistent across datasets | Can be challenging to calibrate, and may produce unclear results when interpreting large, complex graphs or fuzzy data | Relationship detection, metadata redundancy elimination, and quality assurance | [74–76] |
| | GenAI | Generates realistic, high-quality synthetic metadata for missing or incomplete fields, filling gaps in datasets, or creating diverse metadata variations for training purposes | Requires large amounts of data to train effectively, may introduce errors or inconsistencies if not properly monitored | Filling missing metadata, generating synthetic metadata, enhancing dataset diversity | [72, 77] |

**Table 5** (continued)

| Modules | Used AI techniques | Advantages using AI techniques | Limitations | Use cases | Refs |
|---|---|---|---|---|---|
| Metadata storage and indexing | GNNs, KGs | Optimizes the storage and retrieval of complex relationships between metadata elements, improving query performance and scalability | Graph databases can be challenging to scale for extremely large datasets, neural networks require large training data | KG storage, semantic metadata storage, linked data management | [34, 78] |
| | ML/DL | Automates the tagging and labelling of metadata, improving metadata indexing and reducing manual intervention | Requires substantial labelled data for training, may miss nuances in non-standard metadata formats | Automatic metadata tagging for documents, multimedia, and web content | [79, 80] |
| | Ontology-based AI, Semantic Web Technologies | Utilizes ontologies and semantic technologies to improve the indexing and search of metadata | Ontology development can be time-consuming, and semantic indexing can be difficult for unstructured or noisy data | Knowledge management, semantic search, linked data systems | [81–83] |
| Metadata Lineage and Governance | GNNs, KGs, AI-based Traceability Systems | Tracks the history and flow of metadata across systems, ensuring that the origin and transformations of metadata are accurately recorded | Requires careful setup and maintenance of traceability systems; large datasets may be difficult to track effectively | Data lineage, regulatory compliance, audit trails | [84–86] |
| | Blockchain, AI-based Provenance Models | Ensures the integrity and traceability of metadata using decentralized ledger technologies or AI-based systems for auditing and tracking data history | Blockchain can be resource-intensive and slow; AI-based models may require specialized configuration and training | Data integrity, data security, regulatory compliance, audit trails | [35, 87] |
| Metadata collaboration and Socialisation | RL | Continuously improves metadata tagging based on real-time user feedback, leading to more accurate and relevant data | Complex to implement, requires substantial interaction data to optimize performance | Collaborative tagging, knowledge sharing, crowdsourced systems | [88] |
| | Ontology-based AI | Ensures consistent metadata and supports collaboration between systems with different taxonomies | Developing and maintaining ontologies can be time-consuming and requires domain expertise | Metadata alignment, knowledge sharing platforms, cross-system metadata integration | [89] |
| | GenAI | Enables collaborative metadata creation across multiple users or systems, maintaining consistency | Requires large datasets and may need manual verification of generated data | Collaborative platforms, metadata creation, knowledge sharing systems | [89] |

**Table 6** Functional Capability Comparison of AI Techniques in Metadata Management

| Technique category | Specific Techniques | Extraction | Classification | Clustering | Reasoning | Real-time support | Human-in-Loop | Explainability | Scalability |
|---|---|---|---|---|---|---|---|---|---|
| Language-Based AI | NLP | ✓ | ✓ | ⚠ | ⚠ | ⚠ | ✗ | ⚠ | ⚠ |
|  | NLU | ✓ | ✓ | ✗ | ✓ | ✗ | ✗ | ⚠ | ⚠ |
| Symbolic/Rule-Based Methods | Rule-Based systems | ⚠ | ⚠ | ✗ | ✓ | ✗ | ⚠ | ✓ | ✗ |
|  | Fuzzy Logic | ✗ | ✗ | ✗ | ✓ | ✗ | ✗ | ✓ | ✗ |
| ML/DL | ML | ✓ | ✓ | ✓ | ✗ | ✗ | ✗ | ⚠ | ✓ |
|  | DL | ✓ | ✓ | ✓ | ✗ | ⚠ | ✗ | ⚠ | ✓ |
| Knowledge-Centric AI | GNNs | ✓ | ✓ | ✓ | ⚠ | ✗ | ✗ | ⚠ | ⚠ |
|  | KGs | ✓ | ✓ | ✓ | ✓ | ✗ | ✗ | ✓ | ⚠ |
|  | Ontology-based AI | ✓ | ✓ | ✗ | ✓ | ✗ | ✗ | ✓ | ✗ |
|  | Semantic Web | ✓ | ✓ | ✗ | ✓ | ⚠ | ✗ | ✓ | ✗ |
| Adaptive/Reinforcement AI | RL | ✗ | ⚠ | ✗ | ⚠ | ✓ | ✓ | ⚠ | ⚠ |
|  | Evolutionary algorithms | ✗ | ⚠ | ⚠ | ✗ | ✗ | ✗ | ⚠ | ✗ |
|  | Bayesian Optimization | ✗ | ✗ | ✗ | ⚠ | ✗ | ⚠ | ⚠ | ✗ |
| Applied and Supportive AI techniques | Collaborative Filtering | ✗ | ✓ | ✓ | ✗ | ✓ | ✓ | ⚠ | ✓ |
|  | Edge AI | ✓ | ✓ | ✓ | ✗ | ✓ | ⚠ | ⚠ | ✓ |
|  | Federated Learning | ✓ | ✓ | ✓ | ✗ | ⚠ | ⚠ | ⚠ | ✓ |
|  | GenAI | ✓ | ✓ | ✓ | ✓ | ✓ | ✓ | ⚠ | ✓ |
|  | Blockchain and Provenance Models | ✗ | ✗ | ✗ | ✓ | ✗ | ✗ | ✓ | ⚠ |

✓ strong support; ⚠ partial/conditional support; ✗ not applicable/not supported

- **Human-in-the-loop:** This allows real-time human interaction during inference, including feedback, correction, or supervision. It requires users to be actively involved in guiding or validating outputs while the system is running—not just during the design, training, or setup stages.
- **Explainability:** Explainability describes how transparently a technique can communicate the reasons behind its outputs. It includes the ability to trace metadata decisions, highlight contributing features, or follow logical steps, which supports user trust, model validation, and regulatory compliance.
- **Scalability:** Scalability refers to a technique's capacity to maintain consistent performance and accuracy as the size, complexity, or diversity of metadata increases. A scalable approach can efficiently handle growth in data volume, input sources, or concurrent users without degradation in response time or system stability.

To further support interpretation, we also group the techniques into broader categories based on their underlying approach. Each category tends to align with certain strengths or limitations across the functional dimensions. A summary of these categories and their typical functional focus is provided below to aid in understanding the broader patterns observed in the Table 6.

**Language-based AI techniques**, such as NLP and NLU, are particularly strong in extraction and classification, making them well suited for tasks involving unstructured text. However, they offer limited support for reasoning or clustering unless integrated with other structures such as ontologies or knowledge graphs. Real-time support is possible with lightweight models. For human-in-the-loop interaction, these techniques do not natively support direct user feedback during inference. While users may review or edit outputs after generation (e.g., validating extracted tags), the system does not adapt its behaviour based on interactive human input unless explicitly integrated with a feedback mechanism or editing layer. NLP and NLU techniques are partially scalable because their ability to maintain performance across large-scale MMS depends heavily on the model size, deployment strategy, and task complexity.

**Symbolic and Rule-Based Methods** show strong performance in reasoning and explainability. These approaches operate using transparent, deterministic logic which enabling clear, traceable decisions based on predefined rules or fuzzy sets. However, the applicability of symbolic methods remains narrow. Both rule-based systems and fuzzy logic provide little meaningful support for metadata extraction, classification, or clustering unless these functions are explicitly and manually defined. This requires significant effort, is difficult to maintain, and does not adapt well to changing data or use cases. Their ability to support real-time processing is also limited, since rule evaluation follows a fixed sequence and is not designed for fast or high-volume operations. While some rule-based systems can allow some level of human involvement through rule editing, supporting interactive feedback during live inference. Another limitation shared by both approaches is scalability. As the number of rules or fuzzy sets increases, the system becomes harder to manage and more likely to suffer from conflicts or inconsistent outputs. Overall, these methods offer strong support for reasoning and explainability, they are not well suited to the dynamic, high-volume, and flexible demands of modern metadata management environments.

**ML and DL techniques** show strong performance in core tasks such as extraction, classification, and clustering. These data-driven methods are widely used in metadata systems for automating the generation of structured metadata from raw inputs. However, both ML and DL lack native reasoning and explainability. They do not explicitly represent logical rules or relationships unless combined with symbolic components. While explainable AI (XAI) techniques have been developed to improve interpretability in DL models. Similarly, human-in-the-loop interaction is not inherently supported. These models operate in a fully automated manner unless extended with external feedback systems or user-facing interfaces that allow dynamic supervision or refinement. ML and DL offer strong scalability. They can be trained on large datasets and deployed efficiently across high-volume metadata environments. When supported by modern infrastructure, these models can handle diverse input types and large-scale workloads while maintaining performance, making them well suited for production-level metadata pipelines that demand adaptability and throughput.

**Knowledge-Centric AI techniques,** including GNNs, KGs, Ontology-based AI, and Semantic Web technologies, consistently support reasoning, enabling systems to infer new relationships, validate logical consistency, or fill in missing information through structured knowledge representations. Their outputs are typically interpretable, especially in symbolic formats like KGs and ontologies, which provide clear and traceable metadata links. These methods generally offer limited real-time support, as graph traversal tend to be computationally intensive. They also lack native human-in-the-loop interaction, as they operate deterministically unless explicitly integrated with a user feedback mechanism. In terms of scalability, GNNs and KGs show partial support where they can handle large knowledge structures but may require significant computational resources or graph simplification techniques to scale effectively. Ontology-based AI and Semantic Web approaches tend to face more severe scalability and performance constraints, especially when deployed in large or dynamic metadata environments.

**Adaptive/Reinforcement AI** including RL, evolutionary algorithms, and Bayesian optimization, are not traditionally designed for direct metadata extraction, classification, or clustering tasks. Instead, their strength lies in adaptive decision-making, policy optimisation, and automated tuning, which can complement metadata workflows when dynamic optimisation or interactive feedback is required. RL stands out slightly in this group, offering support for real-time response and human-in-the-loop interaction, particularly in applications where metadata recommendations or search strategies are refined through user feedback or environment-driven rewards. However, its capabilities in metadata reasoning and explainability are only partial, as RL agents typically learn from trial and error without transparent logic. Evolutionary algorithms and Bayesian optimization are well-suited for optimisation tasks, such as tuning hyperparameters or evolving decision models, but they do not operate on metadata directly and thus score low across most functional dimensions. Their explainability is limited, as outputs result from iterative search rather than interpretable inference. Additionally, scalability is a concern for all three techniques, especially evolutionary algorithms and Bayesian optimization, which can become computationally intensive and slow when applied to large or high-dimensional spaces.

**Applied and Supportive AI techniques** provide auxiliary capabilities that enhance or extend metadata management systems rather than directly performing core metadata tasks. As reflected in the Tables 5 and 6, these techniques show mixed functional coverage. Edge AI and federated Learning both offer strong support for extraction, classification, and real-time processing. Edge AI excels in scenarios where metadata must be processed locally on devices with low latency, while federated Learning allows models to be trained across decentralised datasets without sharing raw metadata, supporting both privacy and scalability. However, both techniques depend on the models they deploy or train, so their support for human-in-the-loop interaction and explainability is often only partial. Collaborative filtering is well-suited for interactive environments where human-in-the-loop feedback plays a role such as user ratings, or preferences directly influencing recommendations. This feedback loop allows the system to refine its outputs over time based on human input, even during live operation. Collaborative filtering can also be implemented with real-time processing, allowing the system to update and refine recommendations immediately as new user interactions occur. GenAI demonstrates strong support across nearly all functional areas in MMS. These models such as large language models, image generators, and multimodal transformers are capable of performing metadata extraction, classification, and clustering directly from unstructured data. They can also support reasoning, particularly through prompt-based logic or few-shot examples, allowing them to generate structured metadata, summaries, or semantic relationships based on context. GenAI techniques are well-suited to real-time processing, especially when optimised for inference using techniques such as model distillation or caching. They also support human-in-the-loop workflows, allowing users to interactively refine prompts, correct outputs, or guide the generation process in metadata tasks. GenAI models can be deployed across large, diverse datasets and integrated into high-volume systems. GenAI provides only partial support for explainability as it naturally does not rely on transparent logic or interpretable rule sets. Instead, it produces outputs based on complex patterns learned from large-scale training data. Although various post-hoc techniques and prompt engineering diagnostics can offer surface-level insights into model behaviour, these explanations are often approximate and may lack consistency. Blockchain and provenance models stand out by providing strong support for reasoning and explainability, especially in the context of metadata integrity, auditability, and governance. Scalability is partially supported, though performance may degrade under high-volume transaction loads.

## 4 Enhancing Metadata Management: Gaps and Solutions

Metadata management has evolved significantly with the integration of AI techniques, offering promising advancements in efficiency, scalability, and accuracy. However, several gaps and limitations persist, both in traditional approaches and AI-powered solutions. This section explores these challenges to identify areas requiring further research and innovation.

### 4.1 Gaps in Current Metadata Management

#### 4.1.1 Unaddressed Challenges in Traditional Approaches

Traditional metadata management approaches, such as rule-based systems and manual processes, often fall short in meeting the demands of modern, large-scale data environments. These methods are increasingly inadequate for handling the scale, complexity, and diversity of contemporary datasets. Key challenges include scalability issues, inflexibility, labour-intensive processes, lack of advanced contextual understanding, and insufficient support for heterogeneous data. These limitations collectively hinder the effectiveness of metadata systems in dynamic, multi-domain applications.

**Scalability issues** arise because traditional systems are unable to manage the vast amounts of metadata generated by big data and diverse sources. These systems are not designed to accommodate the rapid growth in data generation, including unstructured formats and real-time data streams.

**Inflexibility** exacerbates this challenge. Predefined rules and static methods lack the adaptability needed to address the evolving nature of modern datasets, often resulting in metadata that is outdated or irrelevant.

**Labor-intensive processes**, such as manual tagging and validation, further hinder efficiency. These tasks are prone to human error, time-consuming, and impractical for managing large-scale datasets. As a result, maintaining metadata consistency and quality becomes increasingly difficult.

**Lack advanced contextual understanding**, which limits traditional systems' ability to analyse relationships within metadata, interpret semantic structures, and address complex interdependencies. This shortcoming prevents these systems from fully leveraging the potential of intricate datasets.

**Insufficient support for heterogeneous data** presents another critical limitation. Traditional approaches often fail to integrate seamlessly across diverse data types and formats. This is particularly problematic in multi-domain environments where datasets are highly varied. Such limitations reduce the effectiveness of metadata systems, particularly in promoting interoperability and enabling cross-domain analysis.

Because of these limitations, the emergence of AI techniques has become a promising solution to enhance metadata management. AI-powered methods address these gaps by introducing automation, adaptability, and advanced analytics.

#### 4.1.2 Limitations of Existing AI-Powered Solutions

While AI-powered techniques provide significant advantages in metadata management, they also face critical limitations that necessitate the development of more robust and reliable solutions.

**Data Dependency:** AI techniques such as clustering, classification, and generative models rely heavily on large volumes of high-quality labelled data for training. This dependency becomes a significant limitation in contexts where data is scarce, biased, or unrepresentative. Without access to sufficient and diverse datasets, the performance and reliability of these models are compromised.

**Limited Interpretability:** Modern AI models, including DL systems and LLMs, often operate as "black boxes," making it difficult to interpret and validate their decision-making processes. This lack of transparency creates challenges in understanding how results are derived and limits trust in these systems, particularly in critical applications.

**Ambiguity in Metadata Analysis:** AI models frequently struggle with metadata that is vague or contextually nuanced. This challenge leads to inaccuracies in outputs, as models may misinterpret or fail to resolve ambiguous information effectively. Both traditional ML models and LLMs are highly sensitive to the quality and availability of data, which directly impacts their ability to process complex metadata.

**Gaps in the Metadata Lifecycle**: AI-powered tools often fall short across various stages of the metadata lifecycle, including collection, storage, analysis, and maintenance. Maintaining AI systems in dynamic metadata environments requires continuous improvement, involving regular data collection, retraining, and validation. This ongoing process is resource-intensive and can limit the scalability of these solutions.

**Inconsistency and Validation Challenges:** Variability in predictions or outputs across similar metadata sets undermines the reliability and accuracy of AI systems. These inconsistencies make it difficult to ensure uniform performance, reducing the overall dependability of AI in managing metadata.

### 4.1.3 Impact of Gaps on Next-Generation Datasets

The unresolved challenges and limitations in metadata management, as outlined in the preceding sections, significantly affect the ability to effectively manage next-generation datasets. These datasets are characterized by their vast scale, complexity, and diversity, demanding advanced solutions that go beyond traditional and existing AI-powered approaches. Without addressing the identified gaps, the following adverse impacts are likely:

**Reduced Usability and Accessibility:** As noted in above section, existing AI-powered systems often face challenges related to data dependency, ambiguity, and limited interpretability. These issues can significantly limit the usability and accessibility of next-generation datasets. Without accurate and transparent metadata, users struggle to understand and effectively navigate large, diverse datasets, making them less accessible for data-driven decision-making and innovation.

**Loss of Data Value:** The lack of advanced contextual understanding and the dependency on large, high-quality labelled data further exacerbate the problem of data value. Without proper metadata management that can handle complex interdependencies, relationships, and semantic structures, the quality and trustworthiness of metadata degrade. This loss in data value makes it difficult for organizations to leverage their data effectively, reducing the potential for extracting actionable insights from large-scale datasets.

**Inefficiencies in Data-Driven Processes:** Traditional labour-intensive processes, along with gaps in the AI lifecycle as discussed (such as the need for continuous retraining and validation), lead to inefficiencies in metadata management. As datasets grow and evolve, maintaining consistency and quality becomes increasingly difficult. These inefficiencies can slow down the process of data analysis and decision-making, particularly in real-time applications where timely insights are crucial.

**Missed Opportunities for Advanced Analytics:** The limitations of AI models, such as the inability to handle ambiguous metadata or provide clear interpretability, will hinder advanced analytics capabilities. Next-generation datasets require advanced techniques such as pattern recognition, semantic analysis, and predictive modelling to unlock their full potential. However, these methods are often compromised by the gaps in metadata analysis and the inability to manage data effectively, leaving valuable insights untapped.

**Ethical and Regulatory Risks:** The lack of consistency and validation in AI systems also raises significant ethical and regulatory concerns. Without effective metadata management, ensuring data privacy, compliance, and security becomes more difficult, particularly when handling sensitive datasets in fields such as healthcare or finance. These risks are heightened as data governance frameworks struggle to adapt to the complexities introduced by next-generation datasets.

In summary, the limitations and challenges outlined above will significantly hinder the effective management and utilisation of next-generation datasets. Addressing these gaps with more robust, transparent, and scalable AI-powered solutions is crucial for unlocking the full potential of complex and diverse datasets. This approach will help minimize inefficiencies, mitigate ethical risks, and enhance data accessibility, ultimately enabling organizations to derive greater value from their data assets.

## 4.2 Proposed Conceptual AI-Powered Metadata Solution

### 4.2.1 Overview of the Solution

Efficient metadata management has become a critical challenge for organizations seeking to maximize the value of their data assets. A comprehensive solution requires not only the ability to store and organize metadata but also to ensure its quality, governance, and accessibility for informed decision-making. In this paper, we propose a conceptual framework designed to serve as both a reference model and a blueprint for prototyping AI-powered metadata systems for future implementation and evaluation. Rather than presenting a fully operational system, the framework integrates a range of emerging AI technologies and maps their potential roles across key metadata functions. It offers a scalable, adaptable, and secure architecture that can guide the design of metadata management solutions across diverse contexts. The high-level structure of this conceptual framework is illustrated in Fig. 3.

This framework offers a structured approach to the management and utilization of metadata, empowering organizations to unlock the full potential of their data assets. By leveraging cutting-edge technologies, such as ML, DL, NLP, GNNs, KGs and LLM models (e.g., GPTs), it enhances both the depth and breadth of metadata analysis. The framework ensures robust metadata quality, governance, and accessibility, providing tools for automated data validation, classification, and enrichment.

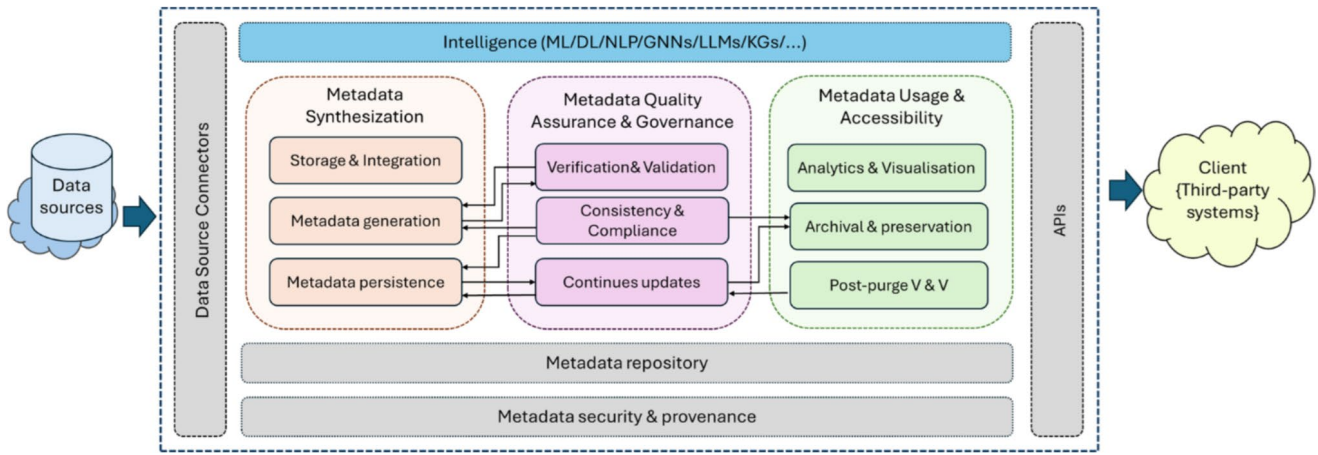

**Fig. 3** High-level architecture diagram

It is designed to be scalable and flexible, supporting integration with third-party systems via seamless API connections. This flexibility allows the framework to be adapted to diverse organizational needs while maintaining a high level of interoperability and usability. The use of advanced ML and DL models ensures continuous improvement of metadata processes, optimizing data workflows and improving decision-making efficiency. Furthermore, its governance capabilities ensure that metadata remains compliant with organizational policies and external regulations, contributing to data trustworthiness and security.

### 4.2.2 Key Features and Capabilities

**Efficient metadata management** is essential for ensuring metadata is accurately collected, generated, and utilized. This feature uses advanced and modern AI technologies to automate processes, enhancing both speed and accuracy. It facilitates seamless integration with external systems, ensuring metadata remains consistent and usable across various platforms. Automation and cutting-edge tools also ensure the system can handle large-scale metadata requirements while supporting long-term retention and scalability.

Key Capabilities:

- **Automated Metadata Generation:** AI technologies automatically create metadata, reducing manual effort and maintaining consistency.
- **Modern AI Integration:** ML, DL, NLP, and generative models are used to process and interpret complex or ambiguous data.
- **Seamless API Integration:** The framework connects with external systems, enabling smooth metadata exchange across different platforms.

- **Long-term Scalability:** It supports the retention and organization of large datasets, ensuring scalability to meet growing demands.
- **Cross-Platform Compatibility:** Metadata remains consistent and interoperable with various tools and systems.

**Quality Assurance & Governance** are critical to maintaining the integrity and reliability of data systems. This feature ensures that metadata is both accurate and compliant with necessary standards, providing the foundation for trust and transparency in any data-driven environment. Through continuous validation and regular updates, the framework adapts to changing data structures, maintaining metadata accuracy and relevance over time. Governance mechanisms also ensure that metadata is protected from inconsistencies and unauthorized changes, preserving its long-term value.

Key Capabilities:

- **Metadata Validation & Verification**: Implements robust verification and validation processes to ensure metadata is accurate, relevant, and trustworthy.
- **Consistency & Compliance**: Aligns metadata with both internal standards and external regulatory requirements, ensuring it meets necessary compliance benchmarks.
- **Data Governance**: Enforces governance policies to maintain metadata integrity, protecting it from inconsistencies and unauthorized modifications.
- **Auditability**: Supports metadata tracking, ensuring that changes and updates to metadata can be traced and audited.

**Advanced Analytics & Accessibility** allows users to gain deeper insights into complex datasets, helping to identify patterns, trends, and correlations that might otherwise go unnoticed. With powerful tools for data analysis and visualization, users can interact with metadata in ways that enhance decision-making and strategic planning. The framework also ensures that metadata is stored and preserved for future reference, maintaining access to critical data as it evolves.

Key Capabilities:

- **Advanced Analytics:** Provides tools to analyse complex metadata, uncovering trends, patterns, and correlations that drive informed decision-making.
- **Data Visualization:** Equipped with powerful data visualization tools to help users interpret metadata and extract actionable insights more intuitively.
- **Metadata Archiving & Preservation**: Ensures long-term preservation of metadata, allowing users to access and reference it in the future, even as it evolves or becomes less frequently used.
- **Insights Generation & Interaction:** Advanced analytics and visualization drive insights that enable better and easy access to metadata, ensuring that users can quickly retrieve and interact with it as needed for analysis or reporting.

### 4.2.3 Potential Impact on Next-Generation Datasets

Next-generation datasets are characterized by their increasing volume, variety, and velocity, posing significant challenges for traditional metadata management systems. The proposed framework is poised to have a transformative impact on next-generation datasets by addressing the challenges of scalability, complexity, and dynamic evolution in modern data environments. By leveraging its advanced features and capabilities, the framework ensures that metadata remains a powerful enabler for data-driven innovation across various domains. Key impacts include:

**AI-Driven Insights for Complex Data Patterns**: Leveraging more advanced modern AI techniques, including generative models (e.g., GPTs), DL models, and NLP, the framework facilitates deep analysis of unstructured, heterogeneous, and high-dimensional data. This enables the identification of hidden patterns, real-time anomaly detection, and predictive modelling, enhancing decision-making precision.

**Interoperability and Ecosystem Integration**: Seamless API-driven integration with third-party platforms ensures interoperability, enabling the unification of diverse datasets and fostering cross-industry collaboration. This capability drives cohesive data ecosystems and reduces operational silos.

**Governance for Trustworthy Metadata**: Robust governance features ensure compliance with stringent industry regulations and organizational standards. Automated validation mechanisms mitigate metadata inconsistencies and vulnerabilities, safeguarding data integrity and enhancing trust in critical applications such as healthcare, finance, and cybersecurity.

**Enhanced Usability Through Analytics and Visualization**: Advanced visualization tools and interactive dashboards provide actionable insights by simplifying the complexity of metadata. This capability supports real-time decision-making in high-stakes domains, including smart cities, autonomous systems, and precision medicine.

**Implementation and Evaluation Outlook**: While the proposed framework is conceptual, it serves as a reference architecture for developing practical metadata management solutions. Its modular design and AI integration offer a flexible foundation for real-world deployment. Future research will focus on evaluating its performance through empirical case studies, assessing the key functional capabilities outlined in Table 6. Demonstrating these in practice will help validate the framework and refine its application in next-generation metadata environments.

## 5 Future Direction and Emerging Trends

MMS must adapt to the fast growth and complexity of modern data environments. This section outlines two key areas. First, it highlights emerging trends in AI and data infrastructure that impact how metadata should be managed. Second, it presents future directions for extending the capabilities of the current framework. These include improving scalability, automation, governance, and system integration. Together, the trends and future developments show how the framework can stay relevant and useful in handling large, changing, and diverse datasets.

### 5.1 Scalable and Sustainable Metadata Infrastructure

#### 5.1.1 Emerging Trends

The increasing volume, variety, and velocity of next-generation datasets demand metadata frameworks that prioritize scalability and adaptability. Future development of the proposed framework should focus on supporting dynamic scalability particularly in domains such as IoT, genomics, and autonomous systems, where data is generated at an exponential rate. Traditional, static metadata systems are often unable to cope with such dynamic and high-throughput environments. Several key trends are emerging to address these challenges:

**Continual Lifelong Learning for Dynamic Metadata:** Continual lifelong learning is gaining traction as a solution for environments where data patterns evolve over time [92]. Unlike traditional batch learning, continual learning enables metadata models to be updated incrementally as new data arrives. This is especially important in streaming contexts such as real-time sensor feeds or biomedical sequencing pipelines where retraining from scratch is not practical. Continual learning supports long-term adaptability, reduces computational cost, and improves the ability of the system to maintain accurate, up-to-date metadata without interruption.

**Cloud-Edge-Device Collaboration:** Cloud–edge–device collaboration enables intelligent distribution of metadata processing across cloud platforms, edge nodes, and endpoint devices [93]. This approach moves beyond traditional cloud-centric models by allowing each layer of the infrastructure to perform specific metadata tasks based on context, resource availability, and latency requirements. Devices can generate or tag metadata locally; edge nodes can perform intermediate processing, validation, or aggregation; and cloud systems handle large-scale storage, analytics, and governance. This multi-tiered collaboration reduces latency, improves responsiveness, and supports real-time metadata operations in distributed environments. It is especially relevant in scenarios such as autonomous systems and sensor networks, where timely metadata processing is critical, and bandwidth is limited.

**Energy-Efficient AI**: As metadata systems scale, their energy and resource demands grow. There is a growing emphasis on energy-efficient AI models and system designs that can deliver high performance with reduced power consumption. Techniques such as model pruning, quantisation, and hardware-aware neural architecture search are being explored to lower computational overhead [94]. These advancements are essential for sustainable operation, particularly in large-scale or resource-constrained environments, such as edge deployments and embedded systems.

### 5.1.2 Future Directions

To operationalise the trends identified above, future development of the framework will evolve in several keyways:

- It will be optimised for distributed deployment across edge, cloud, and hybrid environments. Metadata processes will be designed as loosely coupled, containerised services that can be flexibly distributed based on system load, latency requirements, and data locality. Implementation will leverage orchestration platforms (e.g., KubeEdge) [95] to enable dynamic allocation of metadata tasks such as validation, tagging, or summarisation across the infrastructure stack. This will support use cases involving geographically distributed sensors, mobile devices, and enterprise systems.
- It will integrate continual lifelong learning modules into the metadata model layer. These will be implemented using replay-based, regularisation-based, or dynamically expandable models to enable adaptive learning without catastrophic forgetting. A focus will be placed on integrating these capabilities with streaming pipelines, allowing the system to update metadata rules, patterns, or predictions in near real time as data changes.
- It will include resource-aware scheduling and energy-optimised inference pipelines to support sustainability goals. Rather than treating energy efficiency as an afterthought, model selection and deployment will be guided by profiling tools that consider device constraints and energy budgets. Techniques such as model quantisation, neural architecture search, and low-power edge inference engines will be explored and integrated during deployment phases.

### 5.2 Advanced Modern AI for Metadata Intelligence

#### 5.2.1 Emerging Trends

Recent advances in artificial intelligence are transforming how metadata is understood, generated, and applied across complex systems. Several emerging techniques now offer enhanced capabilities for dealing with unstructured data, automating metadata creation, and improving responsiveness and trust in metadata workflows.

**Multimodal Metadata Understanding:** Modern AI models are increasingly capable of processing and linking multiple data types such as text, images, audio, and video within a unified framework. This enables more comprehensive metadata descriptions that reflect the full context of heterogeneous content. This is especially useful in areas like digital archives [96] or healthcare [97], where systems must handle varied and complex data types.

**GenAI for Metadata Enrichment:** Advanced generative models, such as GPT-based systems, are increasingly used to enrich metadata by producing summaries, descriptive tags, and field values directly from raw content. These models enable automated tagging and offer contextual insights that can be adapted to specific users or domains [98–100]. By enhancing the completeness and relevance of metadata, they support more accurate information retrieval and contribute to improved decision-making. In addition, their ability to personalise outputs helps create more responsive and user-centred applications across a range of settings.

**LLM-as-Agent and Tool-Use Integration**: The recent emergence of LLM-as-agent frameworks, such as LangChain [101], huggingGPT [102], TapeAgents [103] and AgentSquare [104] has introduced the ability for models to plan actions, interact with APIs, and complete structured workflows. In metadata systems, these agents can carry out tasks like schema mapping, format conversion, metadata validation, and consistency checks, autonomously orchestrating multi-step processes with limited human supervision.

**Real-Time Analytics:** It is becoming a key capability in modern metadata management systems, enabling continuous monitoring and adaptive response. Techniques such as TabPFN [105] for fast probabilistic forecasting and real-time event tracing enable rapid analysis and interpretation. These methods support predictive modelling, anomaly detection, and trend analysis, helping systems respond to changes as they occur. This is especially valuable in smart cities and autonomous systems, where accurate and timely metadata processing guides critical decisions and system adaptation.

### 5.2.2 Future Directions

To respond to these trends, the proposed framework will integrate advanced AI modules that enable intelligent metadata operations across various data types and contexts.

- It will implement generative AI components as modular services for automating metadata summarisation, tagging, and attribute generation. These services will be designed to handle multimodal inputs through transformer-based pipelines. A metadata fusion layer will align outputs across formats, and domain-specific configurations will be managed via prompt templates or lightweight fine-tuning. Quality assurance will be built in through rule-based validation checkpoints.
- It will integrate LLM-based orchestration agents that manage metadata operations through predefined workflow templates. These agents will execute multi-step tasks such as schema alignment, metadata cleaning, and cross-source reconciliation by interfacing with internal APIs, metadata logs, and external tools. Their logic will be containerised for reusability and monitored through a task-tracking interface to ensure traceability and error handling.
- It will embed real-time analytics modules into the metadata pipeline using lightweight streaming frameworks. These modules will host models for forecasting, anomaly detection, and temporal pattern recognition, enabling continuous insight generation as metadata flows through the system.

## 5.3 Metadata Governance, Compliance, and Security

As metadata systems scale in complexity and criticality, new governance strategies are emerging to address challenges in compliance, transparency, and security. Recent developments in AI, distributed ledgers, and access control models are enabling more intelligent, accountable, and resilient approaches to metadata management across diverse organisational contexts.

### 5.3.1 Emerging Trends

**AI-Driven Policy Enforcement:** These techniques enhance the automation of compliance monitoring through machine-readable policies and rule-based engines. Approaches such as reinforcement learning are used for policy optimisation in dynamic environments. Integration with policy-as-code tools like Open Policy Agent (OPA) [106] allows for dynamic and scalable enforcement of governance rules. These tools are increasingly applied to validate metadata retention, access control, and transformation policies in real time across complex systems.

**Explainable and Responsible Governance**: Developing dynamic and flexible governance frameworks is essential for adapting to emerging compliance requirements across a wide range of industries. This is particularly critical in domains like healthcare, where privacy regulations are stringent, and finance, where standards evolve rapidly. Tools like AI Explainability 360 [107] help explain how AI models make decisions. Another tool, Croissant-RAI [108], makes it easier for users to find and use responsible AI (RAI) metadata, no matter which platform the dataset is on.

**Decentralized Metadata Management**: This approach is gaining momentum as data ecosystems become increasingly distributed and collaborative. Blockchain and distributed ledger technologies [87, 109] are being explored to securely record and validate metadata histories. These systems offer tamper resistance, verifiable provenance, and shared control mechanisms, supporting transparent and auditable metadata governance across organisational boundaries.

**Zero trust architecture:** Zero trust principles are beginning to reshape metadata governance by enforcing continuous verification and strict access controls [110]. Under this model, all users, services, and data interactions are treated as untrusted by default. Applied to metadata systems, zero trust enables fine-grained authorisation, limits exposure to internal and external threats, and enhances system resilience against unauthorised access or misuse [111].

### 5.3.2 Future Directions

Future development will focus on strengthening metadata governance through automated policy enforcement, enhanced transparency, and secure, accountable workflows.

- It will integrate policy-aware validation mechanisms that continuously monitor metadata flows against defined governance rules. These checks will be embedded in the processing pipeline to enforce compliance in real time and generate alerts or corrective actions when violations occur.
- It will incorporate explainability features grounded in responsible AI principles to ensure transparency in gov-

ernance operations. This includes generating detailed logs, human-readable summaries, and visual reports that document how rules were applied, how decisions were made, and what transformations were performed. These capabilities will support audits, build user trust, and enhance accountability across metadata workflows.
- It will adopt decentralised audit and tracking mechanisms based on cryptographic proofs and distributed ledger technologies. These mechanisms will provide tamper-resistant, timestamped records of metadata activity, enabling verifiable lineage and transparent audit trails for both internal governance and external regulatory reporting. Combined with zero trust-based identity management, this approach will ensure that access to metadata is tightly controlled and traceable across distributed environments.

## 5.4 Cross-Platform Interoperability and Ecosystem Integration

Effective cross-platform interoperability is essential for linking fragmented data systems, promoting collaboration, and enabling the meaningful reuse of metadata across diverse platforms. As data ecosystems grow more complex and distributed, new approaches are emerging to support more seamless integration and interaction.

### 5.4.1 Emerging Trends

**Data Catalogues and Metadata Standards:** These are becoming central to cross-platform interoperability. Recent developments, such as DCAT 2 [112], extend the foundational vocabulary to support richer metadata profiles across diverse domains. The recent work by Sundaram and Musen [113] also highlights how AI-driven approaches are advancing metadata standardisation, further supporting cross-platform integration and FAIR data principles. These enhancements promote semantic consistency and help ensure lossless interoperability between systems.

**Federated Learning Integration**: Integrating federated learning techniques to facilitate secure and private metadata sharing across multiple organizations [114, 115]. By enabling decentralized learning, this approach ensures that sensitive data remains localized within each organization, reducing privacy risks. At the same time, it allows for collaborative insights and model improvements without the need to share raw data, making it particularly valuable for industries like healthcare and finance, where privacy and data protection are paramount.

**Cross-Domain Knowledge Graphs:** Domain-adaptive and cross-platform knowledge graphs are increasingly used to unify metadata from diverse systems. Tools like AmazonKG [116] support consistent linking and integration across services. Recent advances enable aligned feature representations across modalities in cross-dataset scenarios. These approaches enhance semantic interoperability in complex, multi-source environments.

### 5.4.2 Future Directions

To build on these emerging trends, the proposed framework will incorporate several strategic directions to strengthen cross-platform interoperability and ecosystem integration:

- It will adopt advanced semantic metadata standards and knowledge graph integration to support richer, more consistent metadata alignment across diverse platforms. By incorporating tools for schema alignment, ontology mapping, and entity linking, the system will improve semantic interoperability and enable more meaningful metadata reuse.
- It will implement privacy-preserving collaboration mechanisms such as federated learning, allowing multiple organisations to enrich metadata models without sharing raw data. This will support secure, decentralised metadata training pipelines, particularly valuable in sensitive domains such as health and finance.
- It will expand API-driven ecosystem integration by using open and semantically annotated interfaces (e.g., RESTful, GraphQL) [117] to enable context-aware querying and metadata exchange. This ensures adaptability to emerging technologies and smooth integration into heterogeneous digital infrastructures.

## 5.5 Collaborative Development and Validation

Collaborative development is essential for ensuring that metadata frameworks are not only conceptually robust but also practically effective across diverse settings. Increasingly, ecosystem-wide engagement including researchers, developers, industry partners, and users is shaping how metadata systems are evaluated, improved, and maintained.

### 5.5.1 Emerging Trends

**Shared benchmarking environments**: There is growing interest in shared benchmarking platforms where metadata tools and models can be evaluated against standardised tasks and datasets. These environments support transparent performance comparison, reproducibility, and community validation of advanced AI techniques in metadata management (e.g., FAIR assessment tools [118] MetaBench [119]).

**Living frameworks with feedback loops:** Modern frameworks are being designed as "living systems" that evolve through embedded feedback loops. These include continuous integration and stream processing [120], automated improvement pipelines [121], and user-driven modification [122] that ensure sustained relevance and responsiveness.

**Participatory AI Development:** Human-centred and participatory design methods are increasingly used to co-develop AI-based metadata tools with end-users [123]. This trend supports better alignment with user needs, transparency, and acceptance of AI-driven systems.

### 5.5.2 Future Directions

To support robust development and validation, the framework will focus on the following directions:

- It will establish shared benchmarking and evaluation modules, including tasks, datasets, and metrics tailored to metadata workflows. These will allow consistent comparison of components such as metadata extraction, enrichment, and validation tools.
- It will adopt human-in-the-loop validation by engaging domain experts, administrators, and end-users to provide structured feedback on usability, performance, and policy alignment. This participatory input will be integrated into real-world prototyping and empirical testing with academic and industry partners. Pilot deployments across domains will guide iterative improvements, ensuring the framework remains practical, adaptable, and user informed.
- It will foster open-source contributions and community co-development to ensure transparency and continuous evolution. Public repositories, issue tracking, and documentation will encourage adoption and iterative improvement.

## 6 Conclusion

In conclusion, metadata management remains a cornerstone of data governance, resource discovery, and decision-making, particularly in the face of rapidly expanding and increasingly complex datasets. The integration of AI into metadata management processes has the potential to address many of the challenges faced by traditional systems, particularly in the areas of metadata generation, reuse, and lifecycle management. Through the analysis of both traditional and AI-driven metadata approaches, this paper identifies key gaps in current practices and highlights the transformative potential of advanced AI technologies in improving metadata quality and governance.

The proposed AI-assisted metadata management framework offers a promising solution to these challenges, facilitating automated metadata generation, enhancing data governance, and improving the accessibility and usability of datasets. This framework not only supports the effective management of current datasets but also positions organizations to better handle the future complexities of big data.

Looking ahead, further research is essential to refine AI applications in metadata management, with a focus on developing scalable and interoperable systems that can adapt to emerging data types and domains. By embracing AI-driven innovations, organizations can unlock the full potential of their data, driving more informed decision-making and fostering greater resource discovery and reuse.

**Acknowledgements** This work was supported by Australian Antarctic Data Centre (AADC) grant funded by the Australian Antarctic Division. We acknowledge the support and guidance received from the staff at the Australian Antarctic Division throughout the duration of the project. The authors confirm that this work is original and has not been published elsewhere, nor is it currently under consideration for publication elsewhere.

**Author contributions** The authors confirm their contribution to the paper as follows: Study conception and design: Wenli Yang; Muhammad Bilal Amin. Draft manuscript: Wenli Yang; Rui Fu. Data collection and analysis: Rui Fu; Wenli Yang. Supervision: Wenli Yang; Muhammad Bilal Amin; Byeong Kang. All authors reviewed the results and approved the final version of the manuscript.

## Declarations

**Competing interests** The authors declare that they have no known competing financial interests or personal relationships that could have appeared to influence the work reported in this paper.

## References

1. Ulrich H, et al. Understanding the nature of metadata: systematic review. J Med Internet Res. 2022;24(1):e25440.
2. Leipzig J, et al. The role of metadata in reproducible computational research. Patterns. 2021;2(9):100322.
3. Řezník T, et al. Improving the documentation and findability of data services and repositories: a review of (meta) data management approaches. Comput Geosci. 2022;169:105194.
4. Greenberg J, et al. Metadata as data intelligence. Data Intell. 2023;5(1):1–5.
5. Ansari K, Ghasemaghaei M. Big data analytics capability and firm performance: meta-analysis. J Comput Inform Syst. 2023;63(6):1477–94.

6. Chowdhury RH. Big data analytics in the field of multifaceted analyses: a study on "health care management." World J Adv Res Rev. 2024;22(3):2165–72.
7. Naeem M, et al. Trends and future perspective challenges in big data. In: Proceeding of the sixth Euro-China conference on intelligent data analysis and applications. 2022. p. 309–25.
8. Goedegebuure A, et al. Data mesh: a systematic gray literature review. ACM Comput Surv. 2024;57(1):1–36.
9. Aldoseri A, Al-Khalifa KN, Hamouda AM. Re-thinking data strategy and integration for artificial intelligence: concepts, opportunities, and challenges. Appl Sci. 2023;13(12):7082.
10. Liu C, et al. A review of the state of the art of data quality in healthcare. J Glob Inform Manag. 2023;31(1):1–18.
11. Jahnke N, Otto B. Data catalogs in the enterprise: applications and integration. Datenbank-Spektrum. 2023;23(2):89–96.
12. Subaveerapandiyan A. Application of artificial intelligence (AI) in libraries and its impact on library operations review. 2023. 10.6084/m9.figshare.22573345.v1.
13. Kern CJ, Schäffer T, Stelzer D. Towards augmenting metadata management by machine learning. In: INFORMATIK 2021. 2021. p. 1467–76.
14. Oyighan D, et al. The role of AI in transforming metadata management: insights on challenges, opportunities, and emerging trends. Asian J Inform Sci Technol. 2024;14(2):20–6.
15. Lyko K, Nitzschke M, Ngonga Ngomo A-C. Big data acquisition. New horizons for a data-driven economy: a roadmap for usage and exploitation of big data in Europe. 2016: p. 39–61.
16. Mahdavi M, et al. Towards automated data cleaning workflows. Mach Learn. 2019;15:16.
17. Pepper J, et al. Metadata verification: a workflow for computational archival science. In: 2022 IEEE international conference on Big Data (Big Data). 2022. p. 2565–71.
18. Mosha NF, Ngulube P. Metadata standard for continuous preservation, discovery, and reuse of research data in repositories by higher education institutions: a systematic review. Information. 2023;14(8):427.
19. Gartner R, L'Hours H, Young G. Metadata for digital libraries: state of the art and future directions. Bristol, UK: JISC; 2008.
20. Miller SJ. Metadata for digital collections. American Library Association; 2022.
21. Ullah I, Khusro S, Ahmad I. Improving social book search using structure semantics, bibliographic descriptions and social metadata. Multimedia Tools Appl. 2021;80(4):5131–72.
22. Gauglitz JM, et al. Enhancing untargeted metabolomics using metadata-based source annotation. Nat Biotechnol. 2022;40(12):1774–9.
23. Drivas I, et al. Content management systems performance and compliance assessment based on a data-driven search engine optimization methodology. Information. 2021;12(7):259.
24. Melo D, Rodrigues IP, Varagnolo D. A strategy for archives metadata representation on CIDOC-CRM and knowledge discovery. Semantic Web. 2023;14(3):553–84.
25. Formenton D, Gracioso LDS. Metadata standards in web archiving technological resources for ensuring the digital preservation of archived websites. RDBCI Revista Digital de Biblioteconomia e Ciência da Informação. 2023;20:e022001.
26. Karnani K, et al. Computational metadata generation methods for biological specimen image collections. Int J Digit Libr. 2024;25(2):157–74.
27. Kim Y, et al. Githru: visual analytics for understanding software development history through git metadata analysis. IEEE Trans Visual Comput Graphics. 2020;27(2):656–66.
28. Van Dinh C, Luu ST. Metadata integration for spam reviews detection on Vietnamese e-commerce websites. Int J Asian Lang Process. 2024;34:245002. https://doi.org/10.1142/S2717554524500024.
29. Badawy R, et al. Metadata concepts for advancing the use of digital health technologies in clinical research. Digital Biomarkers. 2020;3(3):116–32.
30. Montori F, et al. A metadata-assisted cascading ensemble classification framework for automatic annotation of open IoT data. IEEE Internet Things J. 2023;10(15):13401–13.
31. Boukraa D, Bala M, Rizzi S. Metadata management in data lake environments: a survey. J Libr Metadata. 2024;24(4):215–74.
32. Bhat P, Malaganve P. Metadata based classification techniques for knowledge discovery from facebook multimedia database. Int J Intell Syst Appl. 2021;13(4):38.
33. Elouataoui W, El Alaoui I, Gahi Y. Metadata quality in the era of big data and unstructured content. In: Advances in information, communication and cybersecurity: proceedings of ICI2C'21. 2022. p. 110–21.
34. Paul AK, et al. Efficient metadata indexing for hpc storage systems. In: 20th IEEE/ACM international symposium on cluster, cloud and internet computing (CCGRID). 2020. p. 162–71.
35. Kaur A, et al. Literature review on metadata governance. Open Int J Inform. 2023;11(1):114–20.
36. Stöhr MR, Günther A, Majeed RW. The collaborative metadata repository (CoMetaR) web app: quantitative and qualitative usability evaluation. JMIR Med Inform. 2021;9(11):e30308.
37. Blaxter T, Britain D. Hands off the metadata!: comparing the use of explicit and background metadata in crowdsourced dialectology. Linguistics Vanguard. 2021;7(s1):20190029.
38. Huang V. Information experiences of organisational newcomers: using public social media for organisational socialisation. Behav Inform Technol. 2023;42(9):1279–93.
39. Paterson III H. Dublin Core's DCMIType 'PhysicalObject' and its use across the open language archives community. In: Proceedings of the 17th annual society of American archivists research forum. 2023.
40. Hilbring D. et al. OData-usage of a REST based API standard in web based environmental information systems. In: EnviroInfo 2022. 2022. p. 53.
41. Beyene WM, Godwin T. Accessible search and the role of metadata. Library Hi Tech. 2018;36(1):2–17.
42. Kuduz N, Salapura S. Building a multitenant data hub system using elastic stack and kafka for uniform data representation. In: 19th international symposium INFOTEH-JAHORINA (INFOTEH). 2020. p. 1–6.
43. Rodrigues D, et al. DataHub and apache atlas: a comparative analysis of data catalog tools. In: CAPSI 2022 Proceedings. 2022. p. 41.
44. Šlibar B. Quality assessment of open datasets metadata. University of Zagreb; 2024.
45. Knapen R, et al. Metadata extraction using semantic and natural language processing techniques. In: iEMSs conference. 2014. p. 48.
46. Azimjonov J, Alikhanov J. Rule based metadata extraction framework from academic articles. 2018. https://doi.org/10.48550/arXiv.1807.09009.
47. Hu W, et al. Cleaning by clustering: methodology for addressing data quality issues in biomedical metadata. BMC Bioinform. 2017;18:1–12.
48. Wang Z, et al. Automatic extraction and cluster analysis of natural disaster metadata based on the unified metadata framework. ISPRS Int J Geo Inf. 2024;13(6):201.
49. Rezqa EY, Baraka RS. Document classification based on metadata and keywords extraction. In: Palestinian international conference on information and communication technology (PICICT). 2021. p. 18–24.
50. Ahmed MW, Afzal MT. FLAG-PDFe: Features oriented metadata extraction framework for scientific publications. IEEE Access. 2020;8:99458–69.

51. Jung W, et al. Suicidality detection on social media using metadata and text feature extraction and machine learning. Arch Suicide Res. 2023;27(1):13–28.
52. Choudhury MH, et al. Automatic metadata extraction incorporating visual features from scanned electronic theses and dissertations. In: ACM/IEEE joint conference on digital libraries (JCDL). 2021. p. 230–33.
53. Park H, Chung Y, Kim J-H. Deep neural networks-based classification methodologies of speech, audio and music, and its integration for audio metadata tagging. J Web Eng. 2023;22(1):1–26.
54. Liu R, et al. Automatic document metadata extraction based on deep networks. In: Natural language processing and Chinese computing: 6th CCF international conference. 2018. p. 305–17.
55. Khan R, et al. CrossDomain recommendation based on MetaData using graph convolution networks. IEEE Access. 2023;11:90724–38.
56. Schilling-Wilhelmi M, et al. From text to insight: large language models for materials science data extraction. arXiv preprint arXiv:2407.16867, 2024.
57. Khan R, et al. Impact of conversational and generative AI systems on libraries: a use case large language model (LLM). Sci Technol Libr. 2024;43(4):319–33.
58. Schilling-Wilhelmi M, et al. From text to insight: large language models for materials science data extraction. 2024. https://doi.org/10.48550/arXiv.2407.16867.
59. Gupta T, et al. MatSciBERT: A materials domain language model for text mining and information extraction. NPJ Comput Mater. 2022;8(1):102.
60. Sodhani S, Zhang A, Pineau J. Multi-task reinforcement learning with context-based representations. In: International conference on machine learning. 2021. p. 9767–79.
61. Tsay J, et al. Extracting enhanced artificial intelligence model metadata from software repositories. Empir Softw Eng. 2022;27(7):176.
62. Rohatgi S. Design and data mining techniques for large-scale scholarly digital libraries and search engines. The Pennsylvania State University; 2023.
63. Nahta R, et al. Embedding metadata using deep collaborative filtering to address the cold start problem for the rating prediction task. Multim Tools Appl. 2021;80:18553–81.
64. Visengeriyeva L, Abedjan Z. Metadata-driven error detection. In: Proceedings of the 30th international conference on scientific and statistical database management. 2018. p. 1–12.
65. Kumar V, Chandrappa, Harinarayana N. Exploring dimensions of metadata quality assessment: a scoping review. J Librarianship Inform Sci. 2024. https://doi.org/10.1177/09610006241239080.
66. Wang Z, et al. A rule-based data quality assessment system for electronic health record data. Appl Clin Inform. 2020;11(04):622–34.
67. Khalid H, Zimányi E. Repairing raw metadata for metadata management. Inf Syst. 2024;122:102344.
68. Tavakoli M, et al. Quality prediction of open educational resources a metadata-based approach. In: IEEE 20th international conference on advanced learning technologies (ICALT). 2020. p. 29–31.
69. Ma A, et al. A deep-learning based citation count prediction model with paper metadata semantic features. Scientometrics. 2021;126(8):6803–23.
70. Quarati A. Open government data: usage trends and metadata quality. J Inf Sci. 2023;49(4):887–910.
71. Ali SJ, Michael Laranjo J, Bork D. A generic and customizable genetic algorithms-based conceptual model modularization framework. In: International conference on enterprise design, operations, and computing. 2023. p. 39–57.
72. Elouataoui W. AI-Driven frameworks for enhancing data quality in big data ecosystems: Error_detection, correction, and metadata integration. 2024. https://doi.org/10.48550/arXiv.2405.03870.
73. Ahmed AA, et al. The role of metadata in promoting explainability and interoperability of AI-based prediction models. J Except Multidiscip Res. 2024;1(1):33–45.
74. Khalid H, Zimanyi E, Wrembel R. Fuzzy metadata strategies for enhanced data integration. In: Proceedings of the 7th international conference on data science, technology and applications. 2018. p. 83–90.
75. Kelley A, Garijo D. A framework for creating knowledge graphs of scientific software metadata. Quant Sci Stud. 2021;2(4):1423–46.
76. Xue B, Zou L. Knowledge graph quality management: a comprehensive survey. IEEE Trans Knowl Data Eng. 2022;35(5):4969–88.
77. Li D, Zhang Z. MetaQA: enhancing human-centered data search using Generative Pre-trained Transformer (GPT) language model and artificial intelligence. PLoS ONE. 2023;18(11):e0293034.
78. Karasikov M, et al. Metagraph: indexing and analysing nucleotide archives at petabase-scale. BioRxiv. 2020. p. 2020. https://doi.org/10.1101/2020.10.01.322164.
79. Wang L, et al. Diesel: a dataset-based distributed storage and caching system for large-scale deep learning training. In: Proceedings of the 49th international conference on parallel processing. 2020. p. 1–11.
80. Sharma A, Kumar S. Machine learning and ontology-based novel semantic document indexing for information retrieval. Comput Ind Eng. 2023;176:108940.
81. Satija M, Bagchi M, Martínez-Ávila D. Metadata management and application. Libr Her. 2020;58(4):84–107.
82. Goy A, et al. Building semantic metadata for historical archives through an ontology-driven user interface. J Comput Cult Herit. 2020;13(3):1–36.
83. Novacek J, et al. Ontology-supported AI model and dataset management. In: IEEE 22nd international conference on industrial informatics (INDIN). 2024. p. 1–6.
84. Aggour KS, et al. Colt: concept lineage tool for data flow metadata capture and analysis. Proc VLDB Endow. 2017;10(12):1790–801.
85. Li M-L. Optimizing data governance through AI-driven metadata management: enhancing data discovery and utilization in organizations. Innovat Eng Sci J. 2022;2(1).
86. Hechler E, Weihrauch M, Wu Y, Intelligent cataloging and metadata management. In: Data fabric and data mesh approaches with AI: a guide to AI-based data cataloging, governance, integration, orchestration, and consumption. Springer; 2023. p. 293–310.
87. Dolhopolov A, Castelltort A, Laurent A. Implementing a blockchain-powered metadata catalog in data mesh architecture. In: International congress on blockchain and applications. 2023. p. 348–60.
88. Behara G, et al. Integrating metadata into deep autoencoder for handling prediction task of collaborative recommender system. Multim Tools Appl. 2024;83(14):42125–47.
89. Mohammed M, Talburt JR, Syed H. Metadata: an integral component of the modern data strategy. In: Congress in computer science, computer engineering, & applied computing (CSCE). 2023. p. 1628–31.
90. Tang C, et al. An empirical case study of meta-IP Chain DAO: the pioneer tokenless DAO. In: IEEE 9th international conference on data science in cyberspace (DSC). 2024. p. 24–31.
91. Tang C, et al. Decentralised autonomous organizations (DAOs): an exploratory survey. Distributed Ledger Technologies: Research and Practice, 2025.

92. Hou J, Cosma G, Finke A. Advancing continual lifelong learning in neural information retrieval: definition, dataset, framework, and empirical evaluation. Inf Sci. 2025;687:121368.
93. Liu X, et al. Brame: hierarchical data management framework for cloud-edge-device collaboration. 2025. https://doi.org/10.48550/arXiv.2502.08331.
94. Theodorou G, Karagiorgou S, Kotronis C. On energy-aware and verifiable benchmarking of big data processing targeting AI pipelines. In: IEEE international conference on Big Data (BigData). 2024. p. 3788–98.
95. Wang S, Hu Y, Wu J. Kubeedge. ai: Ai platform for edge devices. 2020. https://doi.org/10.48550/arXiv.2007.09227.
96. Zhang Z, et al. Multimodal archival data ecosystems. In: IEEE international conference on web services (ICWS). 2024. p. 73–83.
97. Simon BD, et al. The future of multimodal artificial intelligence models for integrating imaging and clinical metadata: a narrative review. Diagnost Intervent Radiol. 2024.
98. Bagchi M. A generative AI-driven metadata modelling approach. 2024. https://doi.org/10.48550/arXiv.2501.04008.
99. Singh M, et al. Leveraging retrieval augmented generative LLMs for automated metadata description generation to enhance data catalogs. 2025. https://doi.org/10.48550/arXiv.2503.09003.
100. Magnus B, et al. Metadata creation and enrichment using artificial intelligence at meemoo. J Digit Media Manag. 2025;13(2):110–23.
101. Asyrofi R, et al. Systematic literature review langchain proposed. In: International electronics symposium (IES). 2023. p. 533–7.
102. Shen Y, et al. Hugginggpt: solving AI tasks with chatgpt and its friends in hugging face. Adv Neural Inf Process Syst. 2023;36:38154–80.
103. Bahdanau D, et al. TapeAgents: a holistic framework for agent development and optimization. 2024. https://doi.org/10.48550/arXiv.2412.08445
104. Shang Y, et al. Agentsquare: automatic llm agent search in modular design space. 2024. https://doi.org/10.48550/arXiv.2410.06153.
105. Liu S-Y, Ye H-J. TabPFN Unleashed: a scalable and effective solution to tabular classification problems. 2025. https://doi.org/10.48550/arXiv.2502.02527.
106. Vadisetty R, Polamarasetti A. AI-powered policy management: implementing open policy agent (OPA) with intelligent agents in kubernetes. Cuestiones de Fisioterapia. 2025;54(5):19–27.
107. Arya V, et al. AI explainability 360 toolkit. In: Proceedings of the 3rd ACM India joint international conference on data science & management of data. 2021. p. 376–9.
108. Jain N, et al. A standardized machine-readable dataset documentation format for responsible AI. 2024. https://doi.org/10.48550/arXiv.2407.16883.
109. Hori H, Oguchi M. a study of blockchain-based metadata management and its use for data verification. In: Twelfth international symposium on computing and networking workshops (CANDARW). 2024. p. 63–8.
110. Fernandez EB, Brazhuk A. A critical analysis of zero trust architecture (ZTA). Comput Stand Interfaces. 2024;89:103832.
111. Mensah F. FastMonitor: enhancing data access control with zero-trust architecture. Int J Acad Indust Res Innov. 2024;10:347–51.
112. Albertoni R, et al. The W3C data catalog vocabulary, version 2: rationale, design principles, and uptake. Data Intell. 2024;6(2):457–87.
113. Sundaram SS, Musen MA. Toward total recall: enhancing FAIRness through AI-driven metadata standardization. 2025. https://doi.org/10.48550/arXiv.2504.05307.
114. Peregrina JA, Ortiz G, Zirpins C. Towards a metadata management system for provenance, reproducibility and accountability in federated machine learning. In: European conference on service-oriented and cloud computing. 2022. p. 5–18.
115. Schlegel M, et al. Collaboration management for federated learning. In: IEEE 40th international conference on data engineering workshops (ICDEW). 2024. p. 291–300.
116. Wang Y, et al. Amazon-KG: A knowledge graph enhanced cross-domain recommendation dataset. In: Proceedings of the 47th international ACM SIGIR conference on research and development in information retrieval. 2024. p. 123–30.
117. Quiña-Mera A, et al. Graphql: a systematic mapping study. ACM Comput Surv. 2023;55(10):1–35.
118. Krans N, et al. FAIR assessment tools: evaluating use and performance. NanoImpact. 2022;27:100402.
119. Kipnis A, et al. Metabench—a sparse benchmark to measure general ability in large language models. 2024. https://doi.org/10.48550/arXiv.2407.12844.
120. Underwood M. Continuous metadata in continuous integration, stream processing and enterprise DataOps. Data Intell. 2023;5(1):275–88.
121. Maalej W, et al. On the automated processing of user feedback. In: Handbook on natural language processing for requirements engineering. 2025, Springer. p. 279–308.
122. Milev P. Development of an information system with user-controlled structure and content. Innov Inform Technol Econ Digital. 2024;1:7–12.
123. Parthasarathy A, et al. Participatory approaches in AI development and governance: a principled approach. 2024. https://doi.org/10.48550/arXiv.2407.13100.

**Publisher's Note** Springer Nature remains neutral with regard to jurisdictional claims in published maps and institutional affiliations.